\documentclass[preprint]{aastex}
\begin{document}

\title{Signatures of Recent Asteroid Disruptions in the Formation and
  Evolution of Solar System Dust Bands}

\author{A. J. Espy Kehoe}
\affil{University of Central Florida, Department of Physics}
\affil{4000 Central Florida Blvd, Building 121 PS 430, Orlando, FL 32816-2385}
\email{ashley.kehoe@ucf.edu}
\and

\author{T. J. J. Kehoe}
\affil{Departamento de F\'{i}sica da Universidade de Aveiro and CIDMA}
\affil{Campus de Santiago, 3810-183 Aveiro, Portugal}
\affil{and}
\affil{Florida Space Institute}
\affil{12354 Research Parkway, 
Partnership 1 Building, Suite 214
Orlando, FL 32826-2933}
\and

\author{J. E. Colwell}
\affil{University of Central Florida, Department of Physics}
\affil{4000 Central Florida Blvd, Building 121 PS 430, Orlando, FL 32816-2385}
\and

\author{ S. F. Dermott}
\affil{University of Florida, Department of Astronomy}
\affil{ 211 Bryant Space Science Center, Gainesville, FL 32611-2055}

\begin{abstract}

We have performed detailed dynamical modeling of the structure of a faint dust band observed in 
coadded IRAS data at an ecliptic latitude of 17$^{\circ}$ that convincingly demonstrates 
that it is the result of a relatively recent (significantly less than 1 Ma) disruption of 
an asteroid and is still in the process of forming. We show here that
young dust bands retain information on the size distribution and
cross-sectional area of dust released in the original asteroid disruption, before it is
lost to orbital and collisional decay. We find that the Emilkowalski cluster is the source of this partial band and that the dust released in the disruption would correspond to a regolith layer $\sim$3 m deep on the $\sim$10 km diameter source body's surface. 
The dust in this band is described by a cumulative size-distribution inverse power-law index with a lower bound of 2.1 (implying domination of cross-sectional area by small
particles) for dust particles with diameters ranging 
from a few $\mu$m up to a few cm. The coadded
observations show that the thermal emission of the dust band structure is
dominated by large (mm--cm size) particles. We find that dust particle ejection velocities need to be a few times the escape velocity of the Emilkowalski cluster source body to provide a good fit to the inclination dispersion of the observations. We discuss the implications that such a significant release of material during a disruption has for the temporal evolution of the structure, composition, and magnitude of the zodiacal cloud.

\end{abstract}

\keywords{minor planets, asteroids: general, zodiacal dust}

\section{INTRODUCTION}
There are now known to be at least three cases where large 
asteroidal bodies, tens to hundreds of kilometers in diameter, have been catastrophically disrupted 
as the result of a large-scale impact within the last ten million years (Nesvorn\'{y} et al. 2003; 2008; Dermott et al. 2002). 
The most dramatic evidence for 
the occurrence of such catastrophic collisions is the existence of asteroid families. These are groups 
of asteroids with very similar orbits that are produced by the disruption of a single, large parent 
body (Hirayama 1918). However, the kilometer-sized daughter fragments that make up the family merely represent the 
large end of a size distribution of collisional debris that extends down to micron-sized dust particles. 
These dust particles can also be observed, in the form of the IRAS (InfraRed Astronomical Satellite, which performed an all-sky survey at infrared wavelengths) solar system dust bands (Low 
et al., 1984) as a fine structure feature of the zodiacal cloud.\\

The three prominent and well studied dust band pairs known to
have been created in asteroidal disruptions within the last 10 My
are associated with the Veritas, Karin, and Beagle asteroid families (Nesvorn\'{y} et al. 2003;
2008).  Several additional dust bands have been
postulated by Sykes (1988) in the IRAS data and by Reach et
al. (1997) in the COBE (COsmic Background Explorer) data but the available observations lacked the sensitivity to characterize the structures.  
The discovery of several new, younger (\textless 1 Ma) asteroid
clusters (Nesvorn\'{y} et al. 2006b) provides possible new
sources for these postulated dust bands. Clusters are similar to asteroid families, but result from groupings of only a few asteroids.  Potential asteroid clusters can be identified from searching for groupings of bodies in proper-element space, where proper elements are  time-averaged orbital elements from which secular gravitational perturbations have been removed.  Milani and Kne\v{z}evi\'{c} (explained in Milani et al. 2014) have produced a large database of proper elements based on the ever-growing catalog of orbital elements from the Minor Planet Center. Using the Hierarchical Clustering Method (HCM; Zappala et al. 1990),  Nesvorn\'{y} et al. were able to identify 
clusters with only a few, small members. These clusters are the result of recent disruptions (Section 3.3) of
much smaller asteroids ($\leq$10 km in diameter) than those producing
the prominent dust bands, whose parent bodies range from
$\sim$30--140 km. Even though these clusters come from smaller parent bodies, they
are so young that they have not yet lost as much of their
original dust (the dust produced in the disruption that created the
family/cluster) as have their older and larger dust
band parent-body counterparts. While the timescale for dust band pair formation in the central asteroid belt from secular gravitational perturbations is $\sim$10$^6$ years (e.g. Sykes and Greenberg 1986), on shorter timescales, dust particles with diameters smaller than $\sim$1 cm are removed by
Poynting-Robertson (P-R) drag, inter-particle collisions,
and radiation pressure. Thus, young, partial dust bands retain more of the dust and preserve the original size distribution of the population ejected into the cloud following an asteroid disruption. \\

 In order to search for these fainter dust bands, we
 utilized a method to coadd the IRAS data, providing us with an
 increased signal-to-noise view of the dust band structure of the
 zodiacal cloud at thermal wavelengths.  In this method, which is discussed in detail in
 Section \ref{sec:coadd}, we coadd virtually all the pole-to-pole
 intensity scans of the IRAS data set using a procedure for correcting
 the variations due to both the observing geometry of the IRAS
 satellite and the structural variations of the cloud.  The
 application of this coadding method to the data produced a
 significant increase in signal-to-noise and revealed and confirmed the existence of
 an additional solar system dust band at an ecliptic latitude of $\sim$17$^{\circ}$. While this feature was glimpsed in earlier work by both Sykes (1988) and Reach et al. (1997), the data at the time lacked the sensitivity to confirm it as a dust band pair and it was thought to possibly be a comet debris trail.  The quality of this coadded data has allowed us to clearly examine the structure of the 17$^{\circ}$ dust band around the sky, confirm it as a dust band pair, and explain its structure.  The data
revealed that the 17$^{\circ}$ dust band is present at some, but not all, ecliptic longitudes.  As will be explained in Section \ref{sec:evidence}, this suggests that the 17$^{\circ}$ solar system dust band is a young, partial dust band which is still in the process of forming (Espy et al. 2009).\\

\noindent

\section{COADDING THE IRAS DATA} \label{sec:coadd}

In order to search for fainter dust band pairs, we increased the signal-to-noise ratio of the observations through a process that involved coadding the IRAS ZOHF (ZOdiacal History File) scans. We corrected structural variations of the cloud itself and those variations resulting from the different observing geometries undertaken by the IRAS satellite (Jayaraman 1995). We normalized and coadded the data in a way that didn't remove or reduce the faint dust band signal.  There are two main variations that needed to be corrected for. The first is the variation of the magnitude of the separation of the band pair seen in observations taken at different solar elongation angles.  This variation stems from a parallax effect due to observing the bands from different distances. The second variation is the shift in the center of symmetry of the dust band pair as seen in observations taken at different ecliptic longitudes.  This variation results because the dust band midplane is inclined to the ecliptic. The details of these variations and their corrections are explained below.

\subsection{Solar Elongation Variations} 
\begin{figure}[ht]
  \centering
    \includegraphics[width=4.0in, scale=0.5]{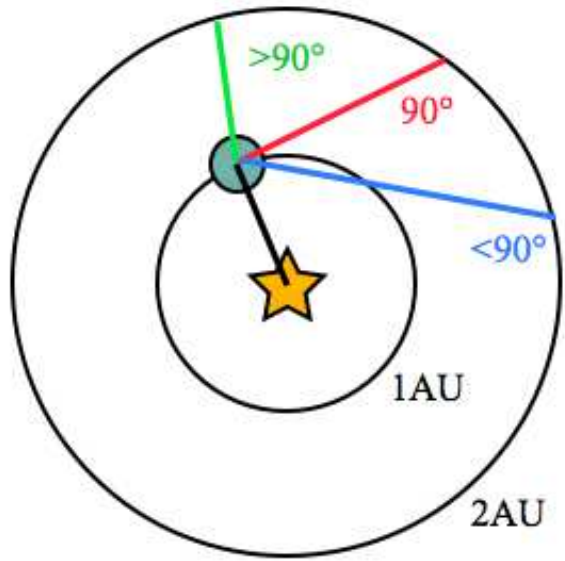}
   \caption[Schematic of solar elongation observing geometry]{Observing geometry variations with solar elongation,$\epsilon$,
 are shown. The Earth (including the observing spacecraft) is shown at 1 AU  (inner circle) and the dust bands are represented as the outer circle at 2 AU.  As IRAS observed at different solar elongation angles, the effective distance to the dust band varied, creating a parallax effect.  Observations taken at $\epsilon>90^\circ$ are closer to the observer than those taken at $\epsilon<90^\circ$.  This results in the effective distance to the dust band varying with elongation angle and explains the variations seen in Figure \ref{fig:elongvar}.}
\label{fig:schem2}
\end{figure}

\begin{figure}[ht]
  \centering
    \includegraphics[width=4in, scale=0.5]{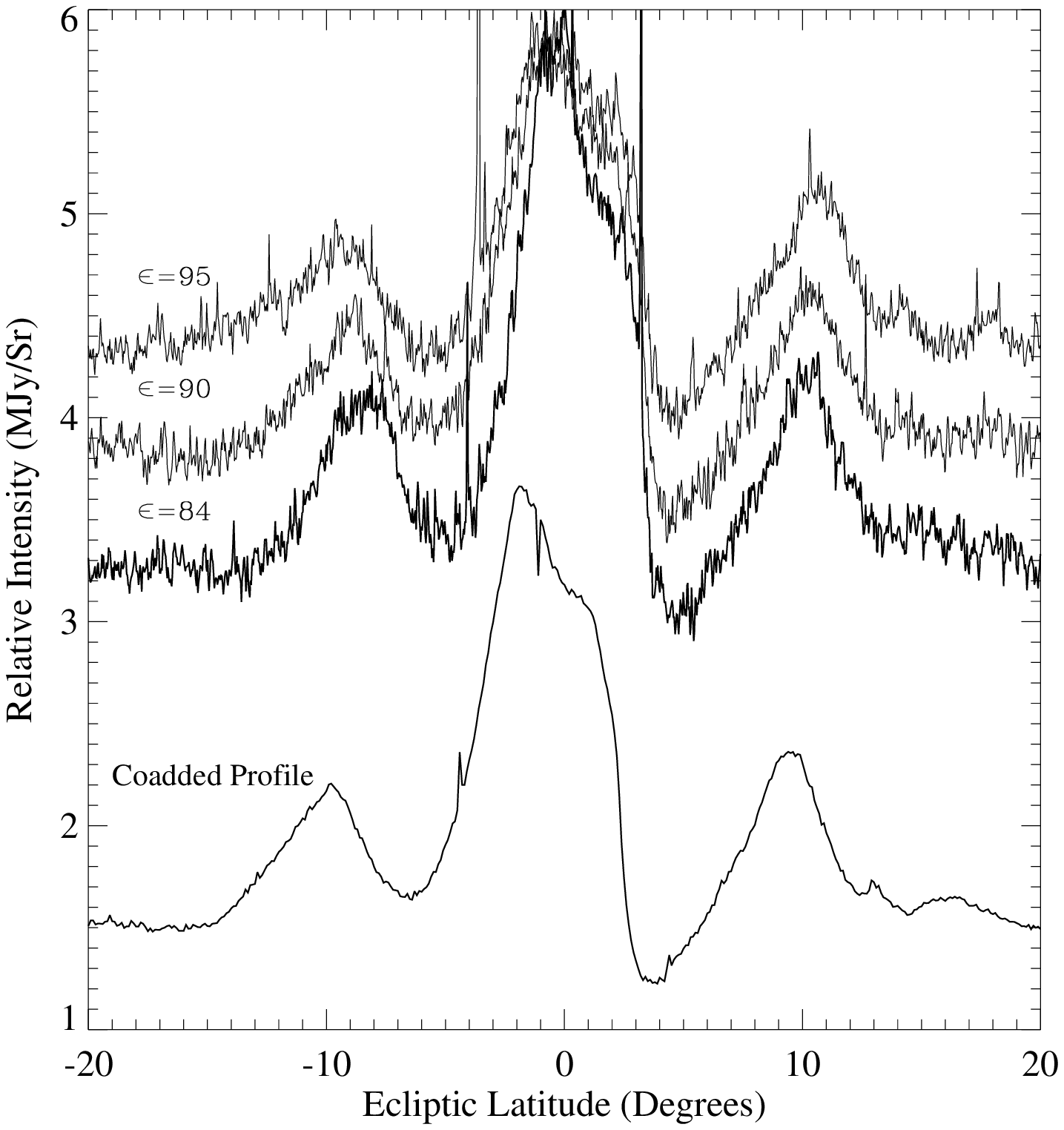}

   \caption[Solar-elongation-induced dust band separation variation]{As IRAS observed at different solar elongation angles, $\epsilon$, the effective distance to the dust band varied, creating a parallax effect.  Observations taken at $\epsilon>90^\circ$ observe dust bands closer to the observer than those taken at $\epsilon<90^\circ$.  This results in a parallax effect of decreasing dust band pair latitudinal separation with increasing distance, as is shown here: a larger separation of the 10$^{\circ}$ bands in the $\epsilon = 95^\circ$ observations and a smaller separation in the $\epsilon = 84^\circ$ observations.  The variation of the separation of the peaks of the north and south 10$^\circ$ bands vary linearly with elongation angle over the small ecliptic longitude range of the lune and this was used to normalize the data to a solar elongation angle $\epsilon = 90^\circ$, so it could be coadded. The coadded data for this range of ecliptic longitude is shown in the bottom profile. }
\label{fig:elongvar}
\end{figure}

In the IRAS observing strategy a lune is defined as a bin of
$\sim$30$^{\circ}$ ecliptic longitude. Each lune contains up to
$\sim$50 usable thermal emission profiles at solar elongation ($\epsilon$) angles varying
from about $\epsilon=85^\circ-95^\circ$.   As the spacecraft incremented the solar
elongation angle of the observations (the observing angle measured
from the Sun-Earth line), the effective distances to the dust bands
changed (Figure 1).  The reason for the difference is that most of the dust band
signal is coming from near 2 AU due to the dynamical dispersion of the
dust band structure interior to this region (e.g. Kehoe et al. 2007).
Observations taken at solar elongation angles greater than 90$^\circ$
result in the spacecraft viewing dust band material that is closer
than the dust band material observed
by the spacecraft when the observations had a solar elongation angle
less than 90$^\circ$ (as shown in 
Figure~\ref{fig:schem2}).  Observing the dust bands at different distances results in a parallax effect: dust band pairs that are closer to the observer (higher elongation angles) will appear to show a greater latitudinal separation than dust that is farther away (smaller elongation angles).  Figure \ref{fig:elongvar} shows an example of the variation in the separation of the 10$^{\circ}$ Veritas dust band pair peaks for observations taken at solar elongation $\epsilon = 84^\circ, 90^\circ$, and $95^\circ$, in a single lune. \\

The 10$^\circ$ Veritas band pair components (seen at an ecliptic latitude of about $\pm$10$^\circ$)  were used as markers of the data for the coadding process.  For each observation, we did systematic fits to determine the precise latitude of the peak flux of the northern and southern components of the 10$^\circ$ band.  When the latitudes were plotted against the solar elongation angle, we found that the location of the latitude of the peak varied approximately linearly with the solar elongation angle of the observation over the ecliptic longitude range of a single lune.  For each lune, we determined the slope and y-intercept of the linear variation then normalized the separation of the peaks in that lune to their $\epsilon = 90^\circ$ values.  An example of a normalized scan is shown at the bottom of Figure \ref{fig:elongvar}.

\subsection{Longitudinal Variations} 
\begin{figure}[ht]
  \centering
    \includegraphics[width=5in, scale=0.5]{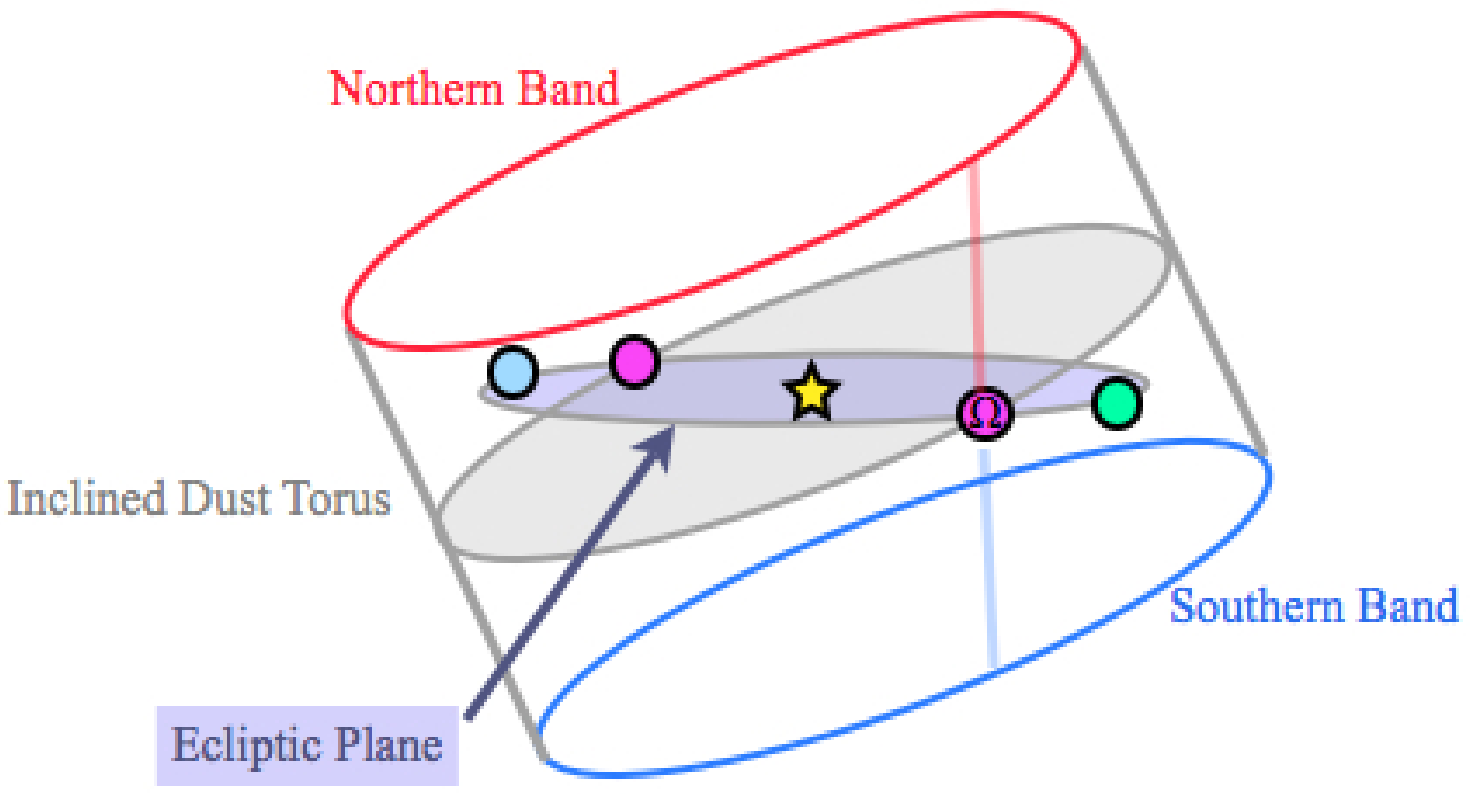}
   \caption[Schematic of an inclined dust torus] {Schematic of an inclined dust torus. The dust torus has a plane of symmetry (the plane about which the dust orbits precess) that is largely determined by Jupiter.  The observing platform follows the Earth and the ecliptic (the Earth's orbital plane), to which Jupiter's orbit is inclined by 1.31$^\circ$.  Because of this inclination, the center of symmetry of the northern and southern bands will vary for observations taken at different ecliptic longitudes.  The northern band is shown in red, the southern band in blue and the intersection of the dust band plane of symmetry (light grey) with the ecliptic (dark grey) is marked at the nodes (purple).
For observations taken of the nodes of the two planes, the northern and southern band will be equally spaced above and below the ecliptic plane (as marked at the ascending node, $\Omega$).  Observations taken between the ascending and descending nodes, (an example is shown in green) will see a shift in the bands with the southern band being closer to the ecliptic than the northern band.  For observations taken between the descending and ascending node (an example is shown in blue) the opposite will be true.  This results in the variation of the band structure shown in Figure \ref{fig:longvar}.}
\label{fig:schem1}
\end{figure}

\begin{figure}[ht]
  \centering
    \includegraphics[width=4in, scale=0.5]{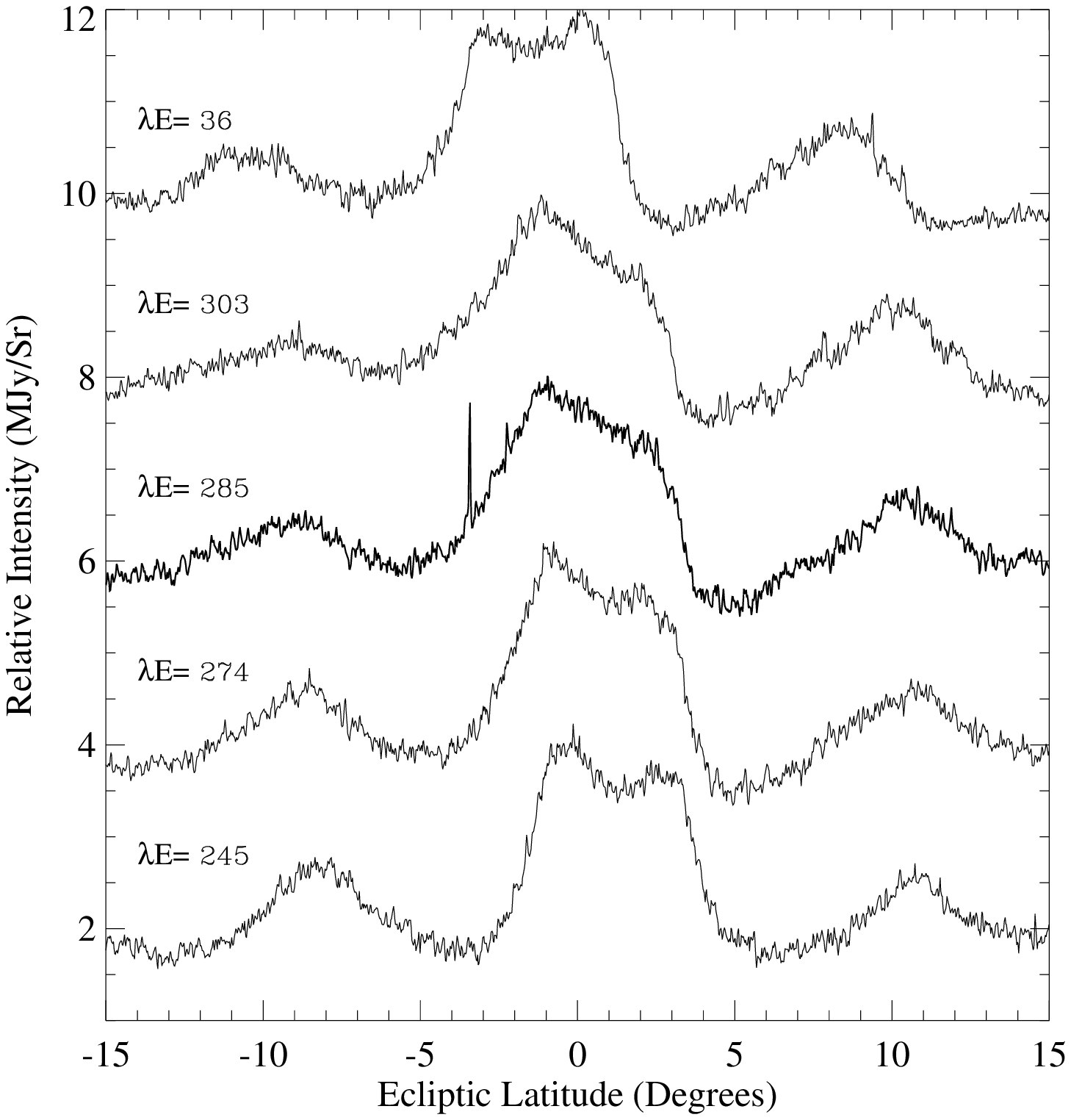}
   \caption[Inclination-induced latitude shift of dust band midplane]{As IRAS observed at different ecliptic longitudes, $\lambda_\mathrm{E}$, the inclination of the dust band midplane with respect to the ecliptic caused an apparent shift of the center of symmetry of the dust band pairs in the different observations.  The variation of the center of the symmetry of the 10$^{\circ}$ dust band pairs is shown for observations taken at five different ecliptic longitudes, each at a solar elongation angle of 90$^\circ$. For observations taken at longitudes away from the nodes of the dust band midplane (where it crosses the ecliptic plane), the midplane of the dust band pairs is not centered on 0$^\circ$ ecliptic latitude.}
\label{fig:longvar}
\end{figure}

Because the dust band midplane is
inclined to the ecliptic (e.g. Dermott et al. 1999), the north and
south components of the dust bands will be equally spaced above and
below the ecliptic only at the nodes of these two planes
(Figure \ref{fig:schem1}). For
observations taken away from the nodes, the bands will have a center
of symmetry that is shifted away from the ecliptic, either to northern
or southern latitudes depending on the ecliptic longitude of the
observation. This effect can be clearly seen if we look at
observations taken at different ecliptic longitudes around the sky.  Figure
\ref{fig:longvar} shows examples for observations taken at a range of ecliptic longitudes (all at $\epsilon = 90^\circ$, representing the approximate center of each lune)) in the direction of the Earth's motion. The seasonal 
shift of the center of symmetry of the 10$^\circ$ band pair due to its
inclination to the ecliptic is clearly evident.  The resulting variation of this shift is sinusoidal around
the sky (e.g. Dermott et al. 1999). Because the inclination of the dust band midplane to the ecliptic is well understood, the magnitude of the shift of the band
center with ecliptic longitude can be characterized and used to normalize the
observations of each lune so that the dust band midplane is centered on an ecliptic latitude of
0$^\circ$.  This removes the forced inclination of the dust band
midplane with respect to the ecliptic (mostly due to the presence of Jupiter) and projects the dust band midplane into the ecliptic plane.  After we normalize the data in each lune to a solar elongation angle of 90$^\circ$, we then normalize these coadded lunes to the ecliptic plane (an ecliptic latitude of 0$^\circ$).  This allows virtually the entire IRAS ZOHF data set to be coadded (with the exception of a few regions contaminated by the galactic center) to yield a high signal-to-noise ratio in the data (Jayaraman 1995; Grogan et al. 1997).

\section{THE 17$^{\circ}$ PARTIAL BAND}\label{sec:seventeen}
\subsection{A New Band Appears}\label{sec:newband}

\begin{figure}[ht]
  \centering
    \includegraphics[width=4in, scale=0.5]{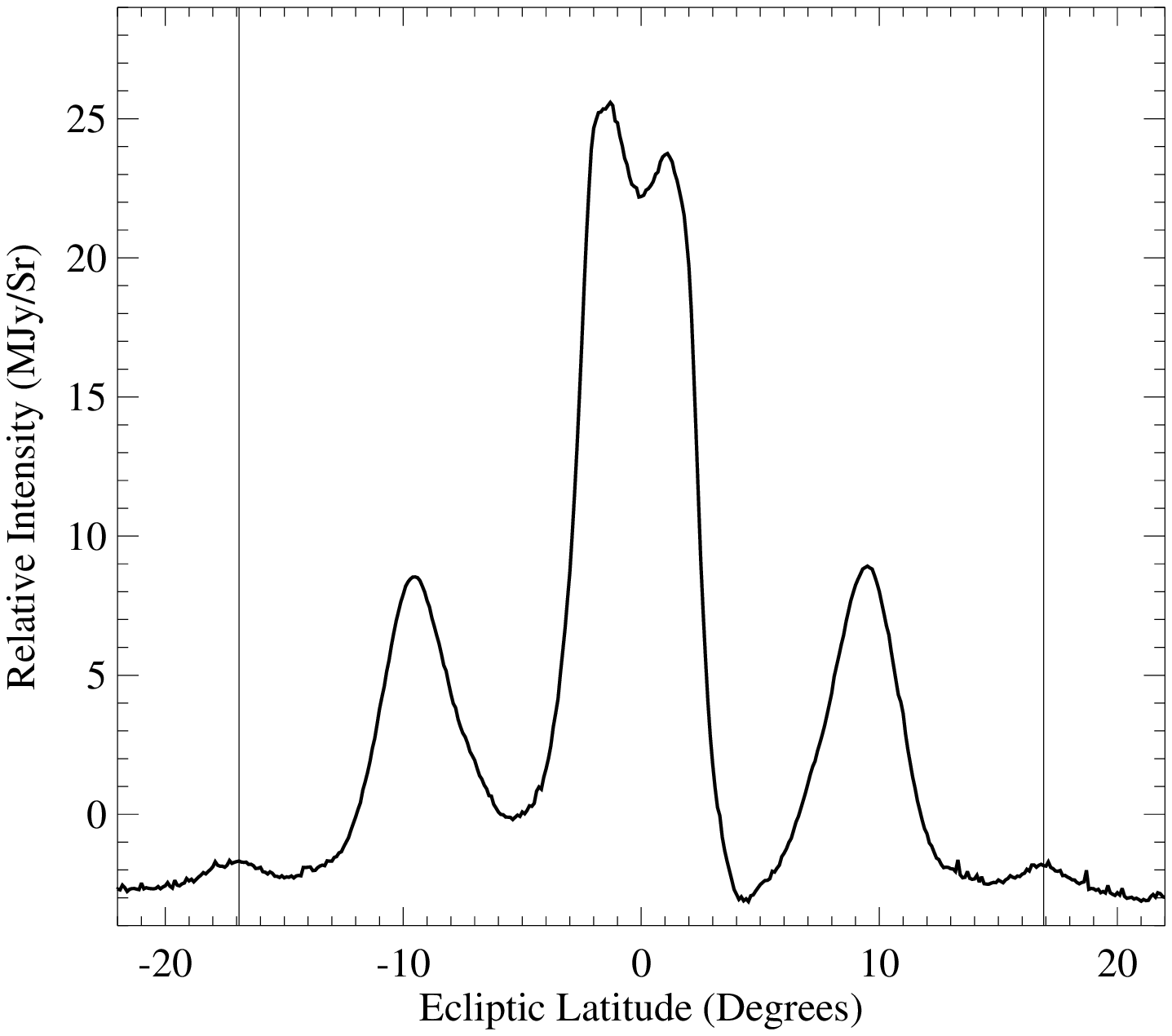}
   \caption[The new dust band as seen in the co-added IRAS data]{The new dust band as seen in the coadded IRAS data in the 25 $\mu$m waveband.  When the Fourier-filtered IRAS ZOHF 25 $\mu$m scans in both leading and trailing directions of the Earth in its orbit and over all ecliptic longitudes are normalized and coadded the new dust band at ecliptic latitude of approximately $\pm17^{\circ}$ can be seen (as marked by the vertical lines).}
\label{fig:alldata}
\end{figure}

We used the coadding procedure described in Section \ref{sec:coadd} to combine virtually the entire IRAS 25 $\mu$m ZOHF data set into a single thermal emission profile. This profile (Figure \ref{fig:alldata})
reveals the existence of a new dust band pair at an ecliptic latitude of approximately $\pm17^\circ$ (shown marked by vertical lines). Based on the latitude of this new band pair, we believe it to be a confirmation of the M/N pair originally seen by Sykes in the IRAS data (1988) and also by Reach in the COBE data (1997).\\

As an additional check that this band is, in fact, a real dynamic structure and not a relic of the Fourier-filtering process (which separates the fine structure dust bands from the broad background cloud), we did a test to determine the plane of symmetry of the new structure relative to the existing band pairs.  We fitted Gaussians to both the northern and southern Veritas $10^\circ$ band components to determine their precise locations and used these to determine the plane of symmetry.  We repeated the process for the northern and southern components of the $17^\circ$ band to determine if both band pairs have the same plane of symmetry, as would be expected if the new dust band is a real dynamical structure. The results of the Gaussian fitting to the partial band and the  $10^\circ$ band show that the partial band does, in fact, have a common plane of symmetry to the other dust band pair to within 0.01$^\circ$, which would be surprising if the band was simply a relic of the Fourier-filtering process.  A common plane of symmetry is expected for all band pairs because, in the asteroid belt where the dust band structure is located, the plane of symmetry is largely determined by the forced inclination imposed by Jupiter (e.g. Dermott et al. 2001).  The inclination of the dust band midplane with respect to the ecliptic was found to be 1.16 $\pm$ 0.09$^{\circ}$ for the central and 10$^{\circ}$ inclination band pairs by Grogan et al. (2001), consistent with the values we determined for the partial band. These values are close to that of Jupiter's orbit which has an inclination of 1.31$^{\circ}$ for the same epoch.  

\subsection{Evidence for a Partial Band}\label{sec:evidence}

\begin{figure}[ht]
  \centering
    \includegraphics[width=4in, scale=0.5]{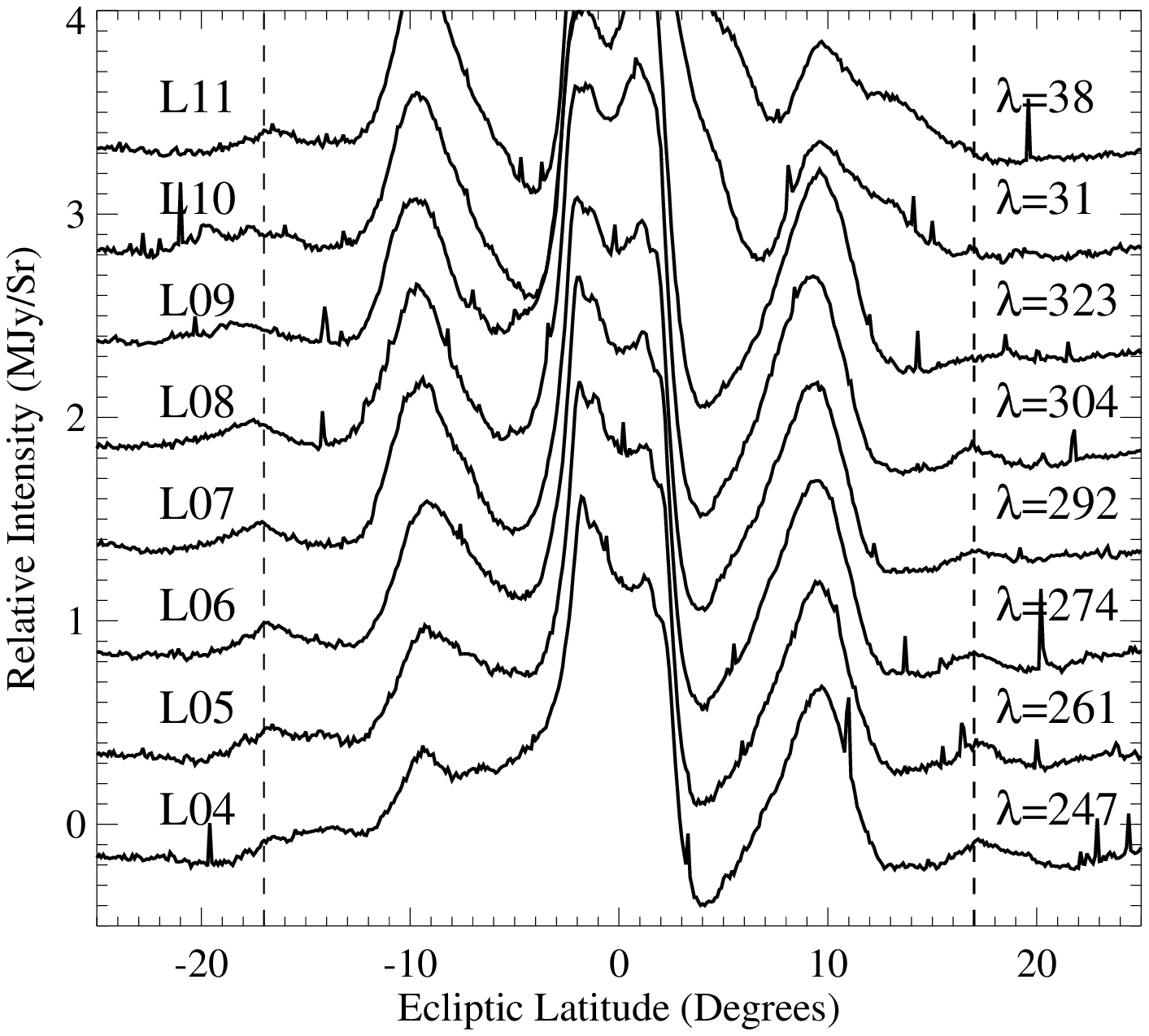}
   \caption[Co-added leading lunes]{Variation of the 17$^{\circ}$ band around the sky can be found
     from the individual lunes (bins of $\sim$$30^\circ$ ecliptic longitude).  This
     figure shows the line-of-sight intensity profiles in the coadded IRAS 25~$\mu$m waveband for all the usable
     (non-galactic-center-contaminated) lunes in the leading direction
     of the Earth's orbit, offset from each other for clarity.  A
     vertical dashed line at $\pm$17$^\circ$ ecliptic latitude marks the
     location of the 17$^\circ$ band.  The lunes are marked with their
     lune designations on the left and the approximate longitude of
     the Earth, $\lambda_\mathrm{E}$,  when the observations were taken. A full dust band pair
     can be seen in lune L08 at 304$^\circ$ longitude but in the L11 lune at 38$^\circ$ longitude, only a southern band can be seen. }
\label{fig:Llunes}
\end{figure}

\begin{figure}[ht]
  \centering
    \includegraphics[width=4in, scale=0.5]{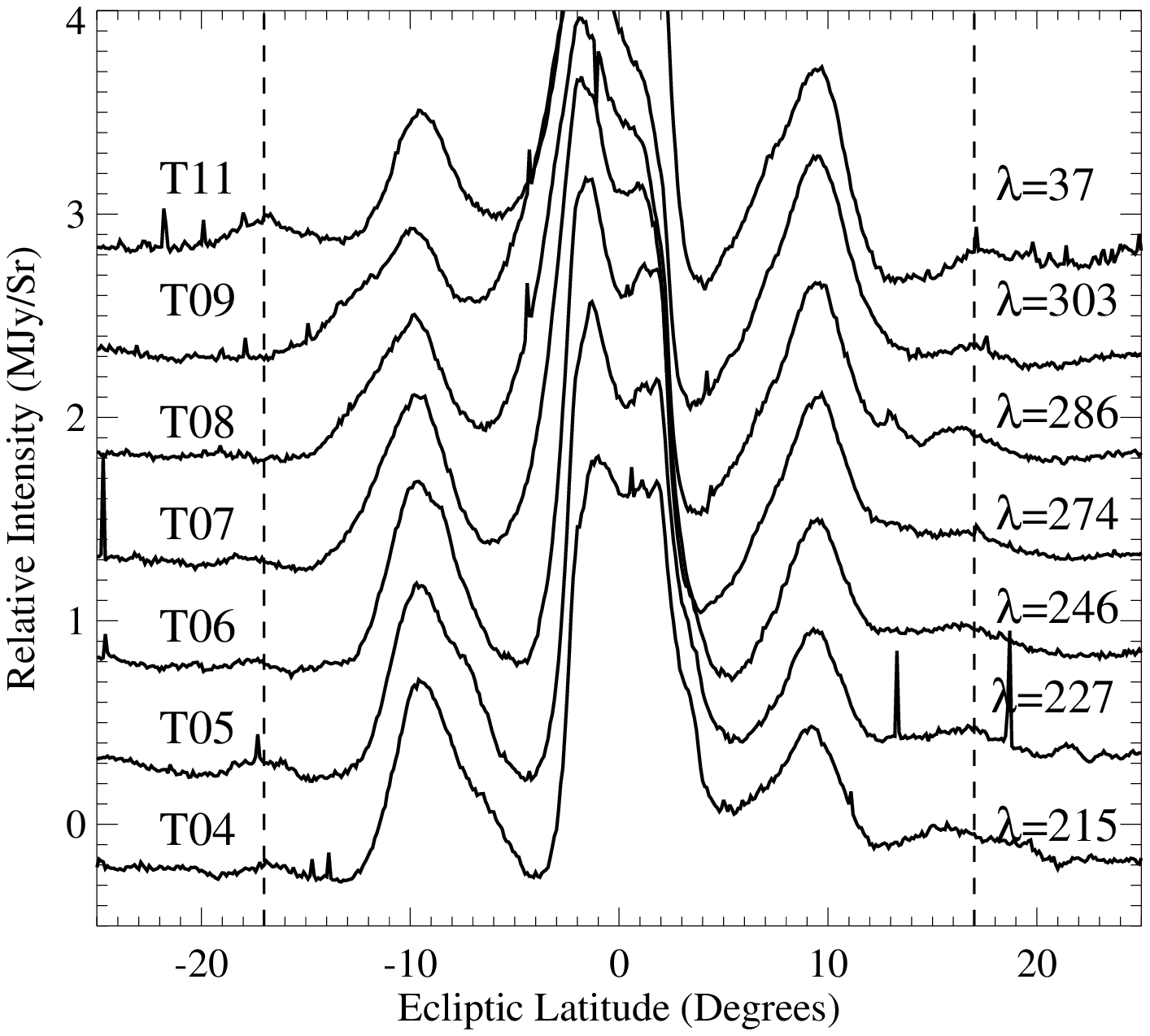}
   \caption[Co-added trailing lunes]{Variation of the 17$^{\circ}$ band around the sky can be found
     from the individual lunes (bins of $\sim$$30^\circ$ ecliptic longitude).  This
     figure shows the line-of-sight intensity profiles in the coadded IRAS 25~$\mu$m waveband for all the usable
     (non-galactic-center-contaminated) lunes in the trailing direction
     of the Earth's orbit, offset from each other for clarity.  A vertical dashed line at $\pm$17$^\circ$ ecliptic latitude
     marks the location of the 17$^\circ$ band.  The lunes are marked
     with their lune designations on the left and the approximate
     longitude of the Earth, $\lambda_\mathrm{E}$, when the observations were taken. A full
     dust band pair can be seen in lune T11 in Figure \ref{fig:Tlunes}
     at 37$^\circ$ longitude but in the T09 lune at 303$^\circ$ longitude, only a northern band is present.  }
\label{fig:Tlunes}
\end{figure}

While coadding the entire IRAS data set serves to clearly reveal and
confirm the existence of the $17^\circ$ band, we can also view the
band in smaller coadded bins of $\sim$$30^\circ$ longitude around the
sky (called lunes) to examine the longitudinal structure
variations. Examining the band in these individual coadded lunes shows
how the magnitude of the band varies around the sky.  These individual
coadded lunes, shown in Figures \ref{fig:Llunes} and \ref{fig:Tlunes},
are for observations taken in the leading and trailing directions of
the Earth's motion, respectively. The 17$^\circ$ dust band is not
present at all longitudes and the intensity of both the northern and
southern components of the band varies. This behavior of the bands gave the first hint that the band at 17$^\circ$ may be a young structure that is still forming and as such, its nodes are not fully differentially precessed around the sky.  To understand this reasoning we consider the formation process of a dust band. \\

When an asteroid breaks up, the orbits of the particles released in
the disruption begin to disperse in semi-major axis,
resulting in a differential precession of their nodes. This results in
the dust orbits slowly forming a torus and a dust band pair. At the initial stage of this process, the particles ejected in the disruption will have
slightly different velocities and thus slightly different semi-major
axes. Usually the velocities of the particles will be on the order of the escape velocity, which for a 10 km diameter asteroid is about 5 m s$^{-1}$ (and for a 100 km asteroid is about 50 m s$^{-1}$).  These ejection velocities are small compared to the orbital velocity of 15--20 km s$^{-1}$, yet result in the particles spreading into a ring of material along the orbit of the parent asteroid on a timescale of hundreds to thousands of years.  Not only are the particles initially spread slightly in semi-major axis from the initial disruption but they are also spiraling inwards under the effect of P-R drag (Wyatt and Whipple 1950), which causes their semi-major axes to decay over time at a rate roughly inversely proportional to particle size.  \\

When the material has just spread around the orbit into a ring, very
early in the evolution, no dust band is evident, just a ring of dust
along the orbit of the source body. As the nodes begin to precess, the
dust bands will start to appear as just a small amount of material at
two different ecliptic longitudes 180$^\circ$ apart in the sky, one at
a  northern latitude and one at a southern latitude.  Slowly, as the
nodes precess and disperse around the ecliptic, the northern and
southern band will spread in longitude and eventually create a full dust band pair when the nodes are completely differentially
precessed. At the intermediate stages of formation there exists at some longitudes a dust band pair, at some only a northern band, and at some only a southern band.  Fully dispersed nodes and a fully formed dust torus are the situation we see for the Veritas, Karin, and Beagle dust bands, which each have full band pairs (e.g. Espy 2010; Nesvorn\'{y}  et al. 2003; 2006a). However, the new 17$^\circ$ dust band appears to be in an intermediate stage of formation, where the nodes are not completely dispersed and the dust bands do not extend all the way around the sky as both a northern and southern band pair.\\

\begin{figure}[ht]
  \centering
    \includegraphics[width=5in, scale=0.5]{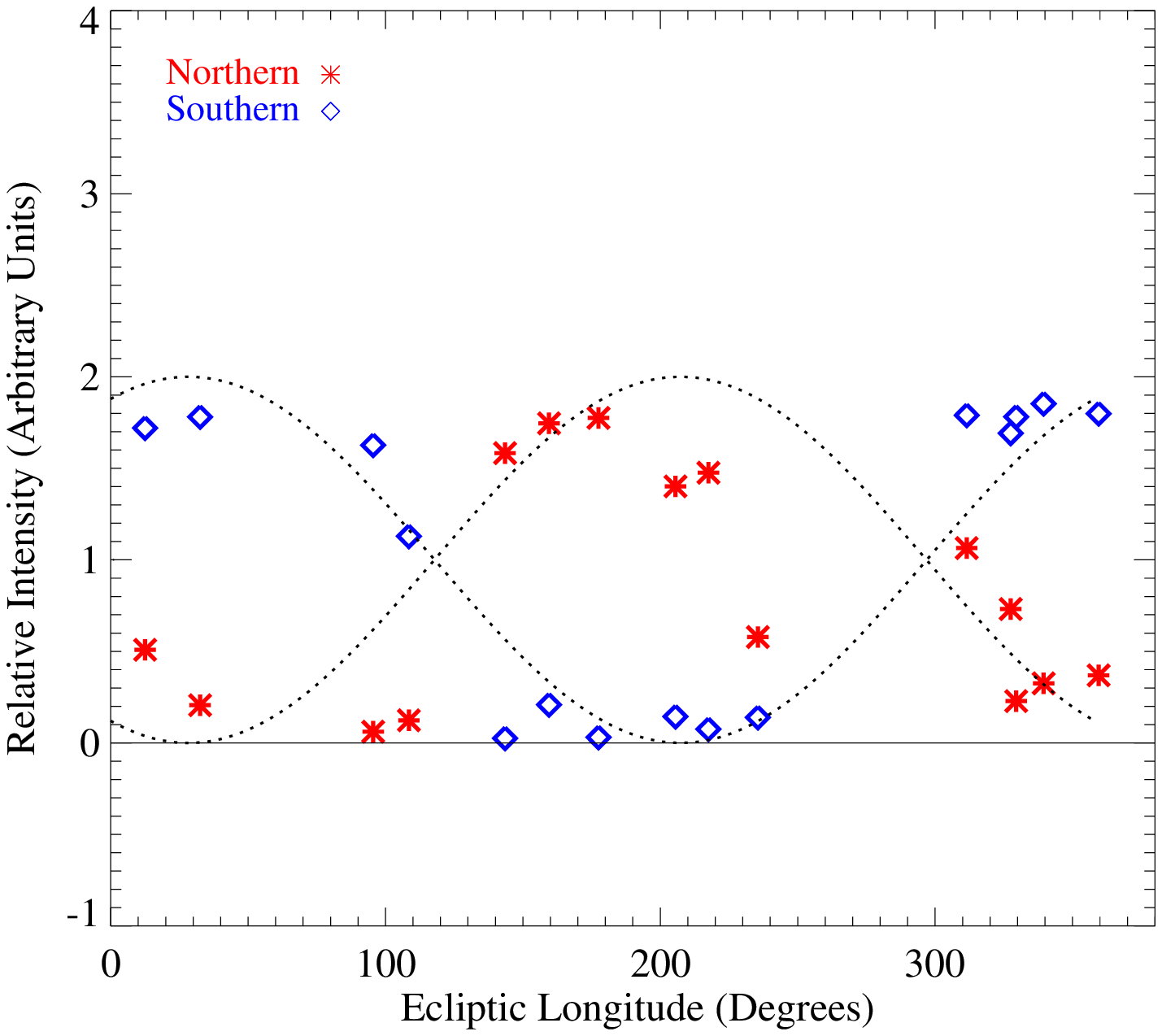}
   \caption[Longitudinal variation of the partial band]{Variation of the 17$ ^{\circ}$ dust band magnitude as a function of ecliptic longitude determined by fitting Gaussians to the northern and southern components of the 17$ ^{\circ}$ dust band in the leading and trailing coadded data (Figures \ref{fig:Llunes} and \ref{fig:Tlunes}). The magnitudes of both the north (red asterisks) and south (blue diamonds) bands appear strong over a range of longitudes and then begin to decay in intensity.  The northern and southern components show this same intensity drop off pattern but 180$^{\circ}$ out of phase, as would be expected for a forming dust band.  A sine curve has been overlaid to guide the eye, but does not represent any actual fit to the data. }
\label{fig:variation}
\end{figure}

We can further investigate the structure of the 17$^\circ$ band by examining how the band intensity varies with longitude. By fitting Gaussians to the northern and southern 17$^\circ$ band profiles in each lune of the leading and trailing coadded observations, we can determine the variation of the intensity of the band emission with longitude. The leading and trailing observations each cover only a portion of the sky. Thus in order to see the band variation around the whole sky we combine these observations by projecting the viewing directions onto the sky. In order to do this, we apply a transformation from the longitude of the Earth when the observations were taken to the longitude of the the structure being observed.  If the dust band structure is considered to be at 2~AU, we can use simple geometric triangulation to determine the longitude variation between the location of the spacecraft and the location of the structures it was observing (in both the leading and trailing directions).  The assumption that most of the flux is coming from 2~AU is not crucial to the result, because if the bands were located in the range of 1.8 AU to 2.5 AU, the longitudinal location of the structure relative to the observations would only shift by a few degrees in either direction.
 The resulting intensity variation of the band is shown in Figure
 \ref{fig:variation}. The bands appear to be bright over a range of
 longitudes, but show a reduced brightness at other longitudes, and the
 northern and southern bands show this pattern 180$^{\circ}$ out of phase.  We overlay a sine curve to guide the eye to the band intensity variation pattern (but it is not a fit to the data).  This pattern of intensity variation is precisely what would be expected by the dynamics of a forming dust band: the northern and southern bands begin to form 180$^{\circ}$ out of phase and spread from these points.  Given some assumptions about the gaps in the data, the northern band appears to begin at about 130$^\circ$ longitude and the southern band appears to begin at about 310$^\circ$ ecliptic longitude.  Based on the way in which the dust band forms, the northern band should be about 90$^\circ$ ahead of the ascending node and the southern band should be about 90$^\circ$ behind it. For the locations of the band observed in the coadded data, 90$^\circ$ behind the start of the northern band would put the ascending node of the source at approximately 40$^\circ$.  Thus, the variation of the dust band components reveals a pattern
that suggests this new dust band is a partial band structure that is still forming and came from a source body with a node of about 40$^{\circ}$.\\

\begin{figure}[ht]
  \centering
    \includegraphics[width=5in, scale=0.5]{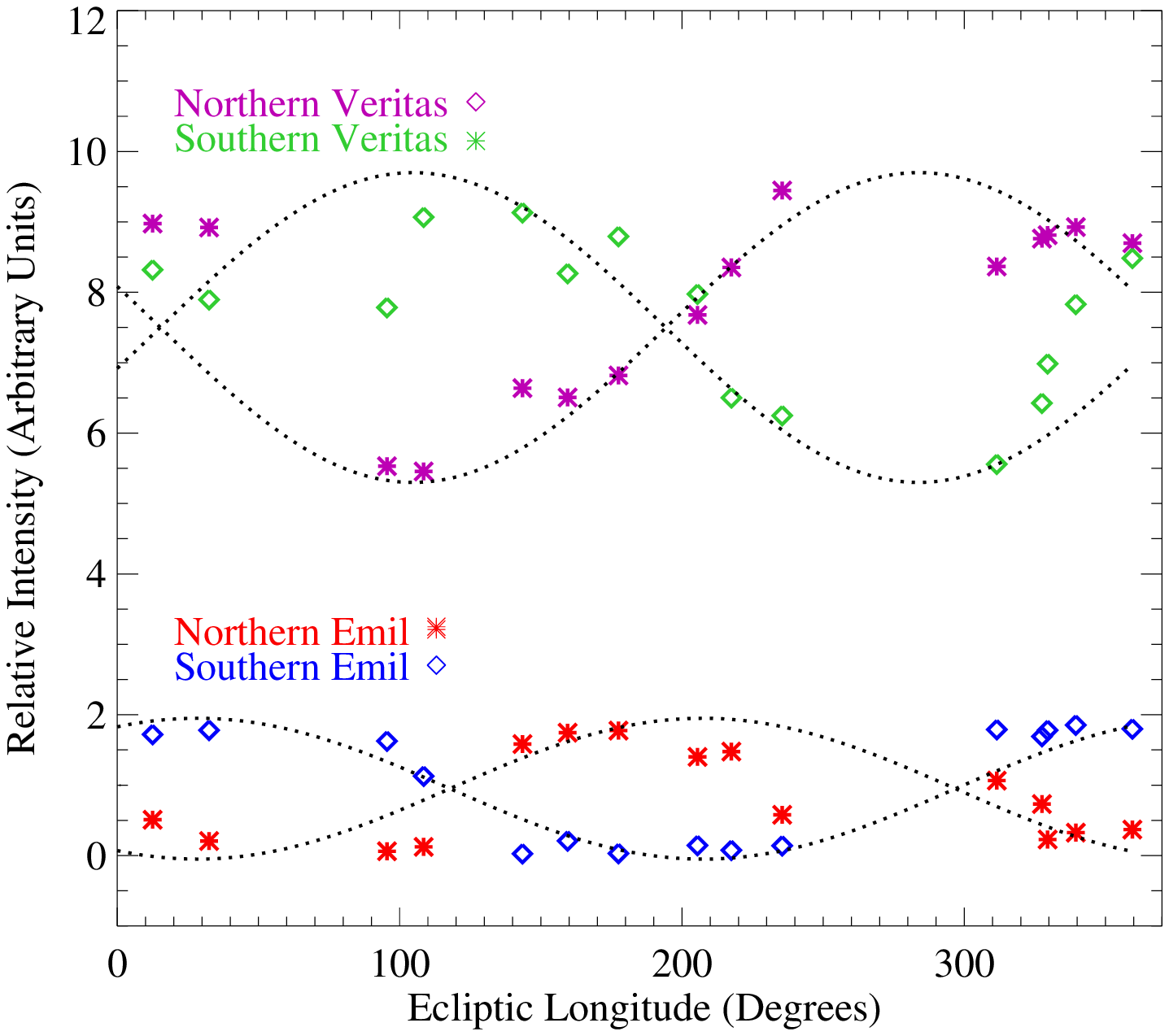}
   \caption[Longitudinal variation comparison with the 10$ ^{\circ}$ band]{Comparison of the longitudinal variation of the 10$ ^{\circ}$ band with the longitudinal variation of the 17$ ^{\circ}$ dust band magnitude determined by fitting Gaussians to the northern and southern components of the 17$ ^{\circ}$ and 10$ ^{\circ}$ dust bands in the leading and trailing coadded data (Figures \ref{fig:Llunes} and \ref{fig:Tlunes}). It can be seen that the intensity variation of the two bands show different phases (different nodes), implying that the variation pattern seen in the bands is not merely a function of a background fluctuation. }
\label{fig:vervariation}
\end{figure}

To be certain that this variation pattern is a function of the new
dust band's structure and not due to a background variation, we also
determined the intensity variation of the northern and southern
components of the 10$^{\circ}$ band peaks using the same method. 
By fitting Gaussians to the northern and southern components of the
10$^{\circ}$ dust bands in the leading and trailing coadded bins
(Figures \ref{fig:Llunes} and \ref{fig:Tlunes}), the band magnitude
variation of the 10$^{\circ}$ Veritas band around the sky can be
compared to the magnitude variation of the 17$^{\circ}$ band.
If the pattern of variation seen in the 17$^{\circ}$ was due to a
background asymmetry, the 10$^{\circ}$ bands would also be expected to
show a similar variation pattern.  Figure \ref{fig:vervariation} shows
the longitudinal intensity variation of the northern and southern
components of the 10$^{\circ}$ band compared to the 17$^{\circ}$ band.
The variation of the 10$ ^{\circ}$ and 17$^{\circ}$ bands shows
variation in a different pattern:  the two bands show different nodes
as well different regions of northern or southern band component
intensity domination. This implies that the longitudinal intensity
variation of the 17$^{\circ}$ band is not merely background cloud fluctuation and therefore provides further evidence that the 17$^{\circ}$ band is a young and still-forming partial band.

\subsection{Possible Sources}\label{sec:possiblesources}
Based on the way in which the dust bands form, the particle orbits
retain the proper orbital inclination of the parent body producing the
band.  This process creates a dust band above and below the midplane
at the latitude of the proper inclination of the parent body and
allows the dust bands to be attributed to possible source bodies in
the main belt.  Using the inclinations of the asteroid families, we can link
the latitude of the new band with asteroid families at that proper
inclination to determine possible sources (e.g. Dermott et al. 1984 ; Nesvorn\'{y} et al. 2003).  At 17$^\circ$ inclination,
the possible sources are 1400 Tirela at a proper inclination of
16.9$^\circ$, and the Emilkowalski cluster at a proper inclination of
17.2$^\circ$.  Because of the intermediate nature of the structure
of this new band, we think it should result from a recent
disruption and, through backward integration of the orbits, the Emilkowalski cluster was found to be 2.2~$\pm$~0.3~$\times$~10$^5$ years old (Nesvorn\'{y} et al. 2006b) and is at a semi-major axis of 2.6 AU, where the timescale for nodal dispersion and thus band formation is on the order of 10$^6$ years.  Thus, we would expect a dust band created from the Emilkowalski disruption to be a partial band, much as we see in the coadded IRAS data. The average node of the Emilkowalski cluster is $\sim$41$^\circ$ degrees (Nesvorn\'{y} et al. 2006b) and consistent with the node needed to fit the data. The other source at 17$^\circ$, the Tirela family, has not been dated because of complications of the backward integrations from chaos due to overlapping mean-motion resonances (Nesvorn\'{y} et al. 2003). As Emilkowalski is the most likely candidate at present, we built a full dynamical model of the dust band that would be associated with the Emilkowalski cluster and compared it to the coadded IRAS data.  Because the three previously known dust bands result from older disruptions, only the inclination remains to determine the source body.  This new band however, allows us more information on the source, because we can model the precession of the nodes of the dust orbits to constrain both the age and the node of the source body. \\

\section{DYNAMICAL MODEL OF A DUST BAND}\label{sec:models}

To produce a model that can be compared with the coadded observations, we track the dynamical evolution of the dust from the likely source body, Emilkowalski.  We use an semi-analytical rather than a purely numerical approach because it allows us to better understand the nature of the early stages of dust band formation and the main forces controlling its formation.  We evolve the dust particle orbits under the effects of  radiation pressure, Poynting-Robertson (P-R) drag, solar wind drag, and perturbations from the 8 planets.

\subsection{Radiation Forces}\label{sec:radiation}

\subsubsection{Radiation Pressure}\label{section:rp}
Radiation pressure is the component of the radiation force that points radially away from the Sun.  It stems from a transfer of momentum due to the impact of solar photons with the dust particles. The strength of the force of radiation pressure varies inversely proportional to the square of the particles' distance from the Sun, just as gravity does. Radiation pressure therefore cancels out some of the Sun's gravitation pull. The motion of a dust particle is the same as the gravitational motion of a particle around an object of mass (1-$\beta$), where $\beta$ is the ratio of the radiation force to the gravitational force (Burns et al. 1979):

\begin{equation}
\label{eq:beta_def} \beta(D)=\frac{F_{\mathrm{rad}}}{F_{\mathrm{grav}}}.
\end{equation}

\noindent
For large spherical particles ($D\geq20\mu\mathrm{m}$) in the solar system, $\beta$ can be approximated (Gustafson 1994): 
\begin{equation}
\label{eq:beta} \beta(D)\approx \frac{1150}{\rho D},
\end{equation}

\noindent
where $\rho$ is the density of the particle in kg m$^{-3}$, and D is the
particle diameter in $\mu$m.   Immediately after the disruption of a
parent body (which itself was too large to be affected by
radiation forces), the dust particles created in the disruption will be perturbed by radiation pressure.  The result is an instantaneous alteration of the orbit of the dust particle calculated by Burns et al. (1979) to be

\begin{equation}
\label{eq:aprime} a^{\prime}=\frac{a(1-\beta)}{[1-2\beta(1+e \cos f)/(1-e^2)]}
\end{equation}

\begin{equation}
\label{eq:eprime} e^{\prime}=\frac{\sqrt{e^2+2\beta e \cos f + \beta^2}} {(1-\beta) }
\end{equation}

\noindent
where $a^{\prime}$ and $e^{\prime}$ are the new semi-major axis and eccentricity of the particle's new, altered orbit, $a$ and $e$ represent the values of the parent body's semi-major axis and eccentricity, and $f$ is the true anomaly which denotes the position in the orbit at the time of disruption.  The orbits of the largest particles, (those with the smallest $\beta$) experience the least perturbation from radiation pressure effects and remain on orbits similar to the orbit of the parent body.  The smaller particles, however, experience a significant orbital perturbation from radiation pressure.  The smallest fragments will end up on hyperbolic orbits (for initially circular orbits, those for which $\beta$$>$0.5; Gustafson, 1994) and thus be removed from the solar system on the order of an orbital period. These are the so called $\beta$-meteoroids.  For asteroidal type densities (2--3~g cm$^{-3}$; Hilton et al. 2002), particles with diameters less than a few microns are the population lost from the system due to radiation pressure, which is referred to as the "blowout threshold".

\subsubsection{Poynting-Robertson Drag}\label{section:PR}
Poynting-Robertson drag is the component of the radiation force that acts tangential to the particle's orbit.  The effect of P-R drag is a decrease in both the semi-major axis and eccentricity of the dust particle's orbit.  P-R drag does not change the plane of the particle's orbit, though, and thus results in no variation of the inclination or the node of the orbit.  The semi-major axis decay caused by the effect of P-R drag is the main transport mechanism of the particles created by asteroid disruptions into the inner solar system. Larger particles decay in more slowly than smaller particles and a useful scaling is that the rate of the semi-major axis decay is roughly inversely proportional to the particle size (see Equation 2). Thus a 200 $\mu$m particle will require ten times as long as a 20 $\mu$m particle to decay into the Sun. The equations for the rate of change of the semi-major axis and eccentricity of the particle orbit were derived by Wyatt \& Whipple (1950):

\begin{equation}
\label{eq:adot} \dot{a}_{\mathrm{PR}}=-\frac{\alpha}{a}\frac{2+3e^2}{(1-e^2)^{3/2}} 
\end{equation}

\begin{equation}
\label{eq:edot}\dot{e}_{\mathrm{PR}}=-\frac{\alpha}{a^2}\frac{2.5e}{(1-e^2)^{1/2}} 
\end{equation}

\noindent
where $a$ and $e$ represent the initial semi-major axis and eccentricity of the particle and $\alpha=6.24\times~10^{-4}~\beta \textrm{AU}^2\textrm{y}^{-1}$.\\

\subsubsection{Solar Wind}\label{section:SW}
Solar wind corpuscular forces are caused by collisions of the dust particles with solar wind protons.  The result on the particles' orbits is a drag effect analogous to that of P-R drag.  The magnitude of the solar wind drag on a particle is estimated to be 30$\%$ of the magnitude of the effect of P-R drag (Gustafson 1994). Solar wind drag, therefore, acts to \textit{increase} the rate of decay of the semi-major axis, decreasing the time it takes for a particle to spiral into the inner solar system.  Solar wind Lorentz forces are ignored for these particles, since this force is significant only for particles less than 1 $\mu$m in diameter in the inner solar system (Gustafson 1994).

\subsection{Ejection Velocity Distribution}\label{section:Vel}
To account for the orbital changes that result from the ejection velocities initially imparted to the dust particles during the asteroid disruption, we use Gauss' equations.  These equations allow the changes in velocity to be converted into changes to the dust particle orbits. The zeroth order form of Gauss' equations (Zappala and Cellino 1996) are:

\begin{equation}
\label{eq:Gaussa} \frac{\delta a}{a}= \frac{2}{na} V_T
\end{equation}

\begin{equation}
\label{eq:gausse}  \delta e= \frac{1}{na} [2(\cos f)V_T + (\sin f) V_R]
\end{equation}

\begin{equation}
\label{eq:gaussi} \delta I=\frac{1}{na} \cos(\omega + f)V_W
\end{equation}\\

\noindent
where $V_T, V_R, V_W$ are the tangential, radial and vertical components of the ejection velocity, respectively, and $n$ is the mean motion.

\subsection{Dynamical Evolution of the Ejected Dust Particles}\label{sec:pbdynamevol}
We calculate the evolution of the dust particle orbits under the effects of secular planetary perturbations from the rate of change of the orbital elements of a particle with time.  These are found from the Lagrange perturbation equations and using the disturbing function and eigenfrequencies from Murray and Dermott (1999). They are given in their final form by:

\begin{equation}
\label{eq:o} \frac{d\Omega}{dt}=\frac{1}{\sqrt{1-e^2}\sin I}\left(BI +\sum_{j=1}^NB_jI_j\cos(\Omega- \Omega_j) \right),   
\end{equation}

\begin{equation}
\label{eq:i} \frac{dI}{dt}=\frac{-\tan(\frac{1}{2}I)}{\sqrt{1-e^2}}e \sum_{j=1}^N A_je_j\sin(\varpi- \varpi_j)+   \frac{1}{\sqrt{1-e^2}\sin I}I \sum_{j=1}^N B_jI_j\sin(\Omega- \Omega_j) ,
\end{equation}

\begin{equation}
\label{eq:e} \frac{de}{dt}=-\sqrt{1-e^2}\sum_{j=1}^N A_je_j\sin(\varpi- \varpi_j),   
\end{equation}

\begin{equation}
\label{eq:p} \frac{d\varpi}{dt}=\frac{\sqrt{1-e^2}}{e} \left(Ae+\sum_{j=1}^N A_je_j\cos(\varpi- \varpi_j)\right) +  \frac{\tan(\frac{1}{2}I)}{\sqrt{1-e^2}}\left(BI+ \sum_{j=1}^N B_jI_j\cos(\varpi-\varpi_j)\right),
\end{equation}\\

\noindent
where the orbits of the planets, denoted by $\Omega_j, \varpi_j, I_j,
e_j$, are calculated using the eigenfrequencies of the system.  These equations are used to
describe evolution with constant $a$, but the
semi-major axes of the asteroidal dust particle orbits are decaying
with time under the effects of P-R and solar wind drag. Combining Equations \ref{eq:adot} and \ref{eq:edot} with
Equations \ref{eq:o}--\ref{eq:p} and expanding the Laplace
coefficients as a series, we can describe the evolution of the dust
particles under both radiative and gravitational forces
simultaneously.  We evaluate these equations over a finite time step
using a fourth-order Runge-Kutta method (Press et al. 1988). The size of the time step is
left as a variable function of particle size due to the different
rates of decay of the dust orbits under P-R drag, and we chose the
time-step to obtain a specific step in semi-major axis, $\Delta a$. We empirically checked the required value of  $\Delta
a$ and chose it to be $\Delta a =10^{-4}$, AU which yields
a time-step of $\sim$10 years for 100 $\mu$m diameter particles and  $\sim$1000 years for 1 cm diameter particles.\\

For each model we use the current orbital elements of the
Emilkowalski cluster (from Nesvorn\'{y} et al. 2006b) and evolve them
backwards in time over the age of the family to determine the orbital
elements at the epoch of family formation. At this epoch we release
the dust particle, using a size range from 10 $\mu$m to 5 cm (a size distribution will be applied later). A velocity is
imparted to each dust particle (to account for the initial impact) as
it is released, and the orbits are altered to reflect this using
Equations~\ref{eq:Gaussa}--\ref{eq:gaussi}.  The ejection velocity of
the particles is allowed to be just slighter greater than the escape
velocity of the parent body ($\sim$10 m s$^{-1}$) as the body is statistically
most likely to be disrupted by a second body that has just enough
energy to do so.  The effect of radiation pressure is then
switched on, yielding another instantaneous orbit alteration
(Equations~\ref{eq:aprime} and \ref{eq:eprime}) dependent on the
particle size. These orbital elements provide the starting point for
the dynamical evolution outlined above. The dust orbits are tracked
into the inner edge of the dust bands near 2 AU. Interior to this semi-major axis, the effect of the secular and mean-motion resonances disperses the dust orbits and
diffuses the band structure into the background cloud (e.g. Espy 2010). The specific semi-major axis location at which each dust particle is removed depends on its size and is characterized by the results of previous numerical simulations (Kehoe et al. 2007). Smaller particles retain their band structure farther into the inner solar system due to their faster P-R drag based orbital decay which results in quicker passage through the resonances and thus reduced perturbation to their orbits. We record the orbital elements of the dust particles at each time and use them to produce the thermal emission models.\\

\subsection{Thermal Emission Models}\label{section:thermal}
In order to compare the results of our dynamical evolution code to the coadded IRAS observations, we need a mechanism by which these orbital distributions can be visualized.  We use the visualization tool SIMUL (Dermott et al. 1988; Grogan et al. 1997) for this purpose.  In short, the SIMUL suite of codes allow the creation of a three-dimensional model of the cross-sectional area of the distribution of dust that would result from a given orbital distribution.  We then visualize the 3-D model as line-of-sight thermal emission profiles which can be compared directly with observational data.  The SIMUL algorithm is based on the idea that a cloud can be represented by a large number of individual dust particle orbits.  The total cross-sectional area of the cloud is divided equally among the orbits. The particles are distributed along the orbits according to Kepler's second law.  The 3-D space is divided into $O(10^7)$ ordered cells, which are filled with cross-sectional area representative of the orbits that pass through the cell.  In this way, the model generates a large three-dimensional array which describes the spatial distribution of the cross-sectional area of dust particle material.  The model is then converted to line-of-sight thermal emission profiles that match the observing geometry of IRAS (or any telescope) by calculating the Sun-Earth distance and ecliptic longitude of Earth at the observing time and setting up appropriate coordinate systems.  These profiles can be compared directly to the IRAS scans in order to constrain the properties of the dust particles composing the bands.\\

\subsection{Particle Size Distribution}\label{section:sizedist}
We implement a size distribution in the models and use the comparison with the coadded data to constrain the parameters of the distribution.  The size-frequency distribution of particles resulting from a catastrophic collision can be described by a cumulative power-law of the form

\begin{equation}
\label{eq:Ncum} N(>D)= \frac{1}{3(q-1)}\left(\frac{D_{0}}{D}\right)^{3(q-1)}
\end{equation}

\noindent
where$ N(D)$ is the number of particles with diameter greater than $D$, $q$ is the size-frequency index, $D$ is the particle diameter, and $D_{0}$ is a constant as 
\begin{equation}
\label{eq:Do} D_0= \sqrt[3]{3(2-q)} D_{\textrm{e}}\left(\frac{D_0}{D_{\mathrm{max}}}\right)^{2-q}
\end{equation}
\noindent
where $D_{\mathrm{max}}$ is the diameter of the largest fragment of the disruption and $D_e$ is the equivalent diameter of the entire distribution.  A distribution in a closed box in collisional equilibrium (Dohnanyi 1969) is described by a power law with $q = 11/6$ and, in this case, the cross-sectional area of the distribution is dominated by particles at the small end of the size range. The value of $q = 5/3$  represents a turnover value where distributions described by a larger $q$ will have area dominated by the smallest particles of the distribution and a distribution described by a smaller $q$ will have area dominated by the largest particles present in the distribution. \\

We implement the size distribution of the particles in the models through the choice of particle sizes making up each of the bins that come together to produce a final dust torus model.  For each size distribution that is to be modeled, the particle sizes are divided into 20 bins. These bins are chosen using a logarithmic distribution of sizes such that there is an equal amount of cross-sectional area in each bin.   As the size distribution parameter, $q$, changes between models, so does the distribution of particle sizes in the bins.  For example, a size distribution dominated by small particles (higher $q$) has more bins of small particles and fewer large particles than would those for a  model with a size distribution described by a smaller value of $q$. This method also allows the cross-sectional area present in the model to be constrained separately from the size distribution since changing the area in this regime would increase (or decrease) the amount of area in the bins together but would not affect the relative contributions of each bin.   These bins are used to create 20 individual SIMUL models, each with the size, node and semi-major axis ranges describing the particles in that bin.  Each SIMUL model is then converted into a thermal emission torus, described by the emission of the average size particle in that bin.  The 20 thermal emission tori are then added together to produce a full model of the partial dust band resulting from the contributions of all particle sizes.  In this way, we can compare observations to models to constrain the size distribution of the dust particles by matching the variation of the band intensity with ecliptic longitude. \\

\section{THE ROLE OF COLLISIONS}\label{sec:collisions}
Given that this new dust band appears to be very young, it is possible that collisions might not yet be playing a dominant role.  If the timescale for collisions is longer than the age of this dust band, then this band represents a structure whose dynamics are dominated mostly by P-R drag and planetary perturbations.  If collisions are not yet affecting the size distribution of the dust band particles or acting to remove large quantities of dust, then we can use the 17$^{\circ}$ band to constrain the amount of dust created in an asteroidal disruption and the size distribution of the dust that is produced in these catastrophic collisions. In order to determine the role of collisions in the early stages of dust band formation, we follow the analysis of Wyatt et al. (1999).\\

Assuming that the size distribution of particles can be described by a single power-law and that all particles have approximately the same density, Wyatt et al. (1999) have shown that the collisional lifetime of a particle of diameter $D$ at a distance $r$ from the Sun can be expressed (as a function of its orbital period, the optical depth of the zodiacal cloud, and some function of the size distribution of particles) using mostly geometric arguments.\\

\noindent

To determine the collisional lifetime we also need to calculate the minimum particle size capable of catastrophically disrupting a particle of size, $D$.  At the relative velocities of particles in the main belt ($\sim$5 km s$^{-1}$
), Dohnanyi (1969) found that the impactor mass needed to overcome the tensile strength and gravitational energy (for the larger bodies) to catastrophically disrupt a body is given by

\begin{equation}
\label{eq:massrat} \frac{M_{\mathrm{imp}}}{M_{\mathrm{body}}} \ge 10^{-4}.
\end{equation}

\noindent
Assuming that the body and the impactor have the same density, this can be converted into a diameter ratio given by 

\begin{equation}
\label{eq:dratio} \frac{D_{\mathrm{imp}}}{D_{\mathrm{body}}} \ge \sqrt[3]{10^{-4}}
\end{equation}

\noindent
or

\begin{equation}
\label{eq:dcc} D_{\mathrm{cc}}(D) \ge 0.046 D.
\end{equation}

\noindent
where $D_{\mathrm{cc}}$ is the diameter of the smallest body capable of disrupting a body of size $D$.
Assuming a value of $q = 11/6$ (to represent an old, collisionally evolved population) for the background cloud particles and that the measured normal optical depth of the cloud at 1 AU, $\tau_{\mathrm{eff}}$ is $O(10^{-7}$), is also representative of the order of magnitude in the  main belt, yields a useful approximation for the collisional lifetime of asteroidal dust particles in the main belt ($\textrm{r} = 2.6$ AU is used), as

\begin{equation}
\label{eq:tcolap} t_{\mathrm{coll}} \approx 2.5 \times 10^{4}\sqrt{D}
\end{equation}

\noindent
where $D$ is the diameter of the particle in microns and $t_{\mathrm{coll}}$ is in
years.  It should be noted that the collisional lifetime of a particle
is commensurate with size, and the larger a particle is, the longer it
will live.  This is due to the nature of the size distribution, as there are less particles in the distribution which are large enough to break up a body the larger that body is.\\  

\begin{figure}[ht]
  \centering
    \includegraphics[width=6.0in, scale=0.5]{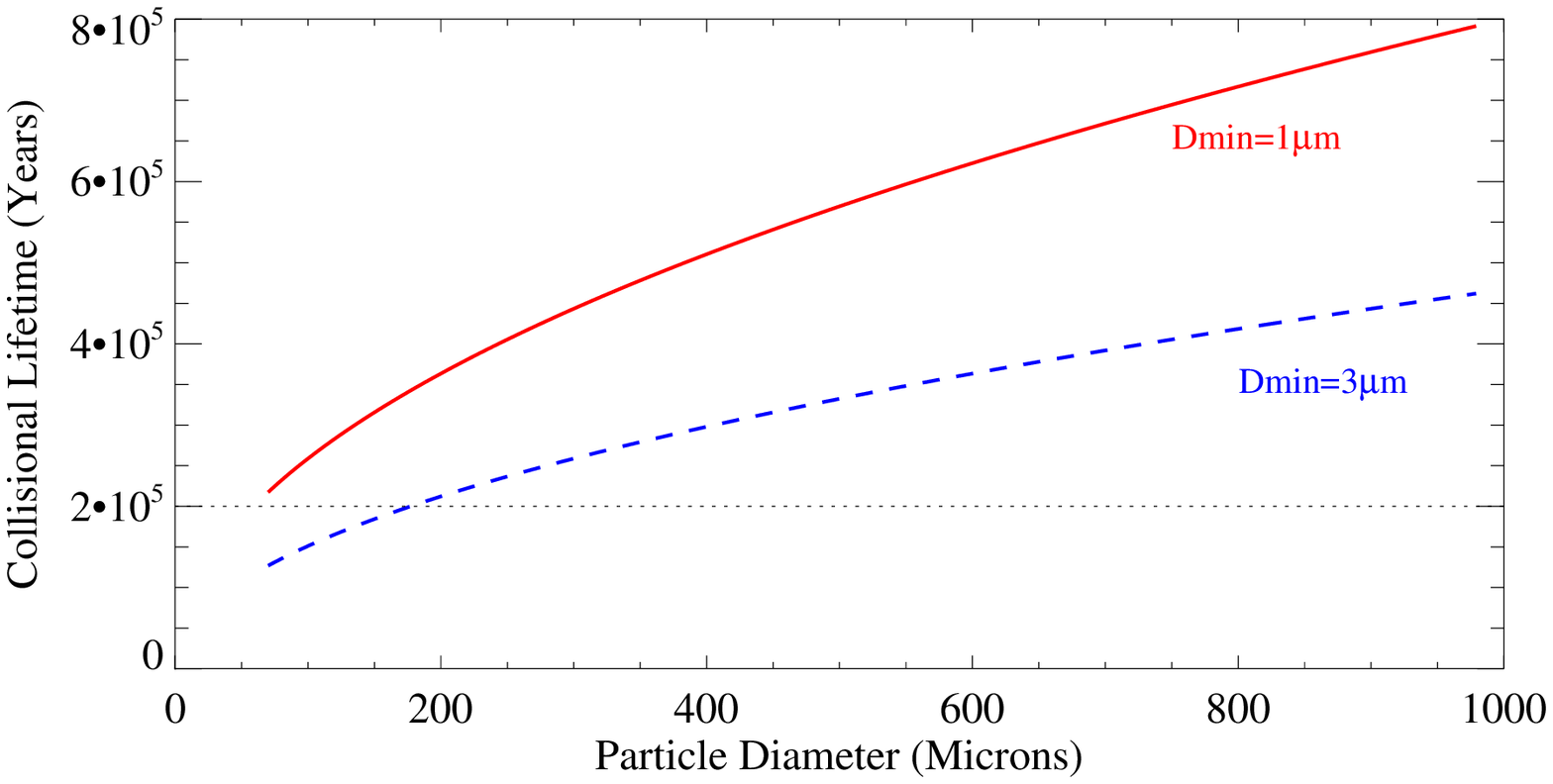}
 
\caption[Collision timescales]{Collisional lifetime as a function of particle size is shown for two different values of the minimum particle size in the background population, determined by the radiation pressure blowout threshold (which is dependent on the density assumed for the particles). Only the collisional lifetime for particles $\geq$~70~$\mu$m are shown, as this is the approximate minimum size particle still contributing to the dust band structure for the 2.2 ka Emilkowalski cluster.  The background population is described by a $q$=11/6 size distribution to represent an old, evolved background population.  The solid line is the collisional lifetime for a particle distribution in which the smallest particle remaining from radiation pressure blowout is 1~$\mu$m (as would be expected for silicates, Backman et al. 1995). The dotted line assumes a blowout threshold of 3 $\mu$m. The dashed line represents the approximate age of the Emilkowalski cluster breakup, and thus particles whose collisional lifetime is above this line have not yet had time to breakup.}
\label{fig:colltime}  
\end{figure}

The results of this analysis are shown in Figure \ref{fig:colltime}. The solid line is the collisional lifetime for the particles in which the smallest particle remaining from radiation pressure blowout is 1 $\mu$m (as would be expected for silicates; Backman et al., 1995), and the dashed line is the collisional lifetime for a minimum particle size of 3 $\mu$m (to represent the larger end of the expected radiation pressure blowout threshold size for a range of compositions). Both distributions assume a background size distribution described by $q = 11/6$.  The dotted line represents the approximate age of the Emilkowalski breakup.  The timescale for collisions is longer for a minimum particle size of 1 $\mu$m because, for the given optical depth of material in the cloud described by a power-law size distribution parameter $q\geq$5/3, most of the area is in particles at the smallest end, where they are too small to break up an average size particle in the disk. For particle sizes whose collisional lifetime is above the current age of the system (dotted horizontal line), collisions have not yet begun to dominate their dynamics. It can be seen that collisions have not yet become dominant for any particles larger than $\sim$~70 $\mu$m if the $D_{\mathrm{min}} = 1 $ $\mu m$, as would be expected for silicates. If $D_{\mathrm{min}} = 3$~$\mu m$, collisions would only be important for particles below about $\sim$150--200 $\mu$m, as the collisional lifetime for these particles is longer than the current age of the Emilkowalski cluster. The particle size range for which collisions have become dominant are likely no longer contributing to the dust band structure because most of these particle sizes have already evolved to inside 2 AU and dispersed into the background cloud. \\

A direct comparison between our collisional model and the model of Gr{\"u}n et al. (1985) is difficult across a broad range of particle sizes because we adopt a simple, single power-law distribution  for the background cloud (the bullets in the collisional process), whereas the Gr{\"u}n et al. model is a composite of different power-laws based on \textit{in situ} data. For the small particles (up to a few hundred $\mu m$ in diameter) that dominate the cross-sectional area, we find that the collisional timescales are in good agreement, with our model predicting ages a few times longer.  Since collisional lifetimes are also a function of orbital period and optical depth  (Wyatt et al. 1999), this is the level of difference we would expect given that Gr{\"u}n et al.'s values are determined at a semi-major axis of $1\,\mathrm{AU}$ and our  values are determined at $2.6\,\mathrm{AU}$, the source region of the Emilkowalski cluster.  However, for larger particles (up to a few cm in diameter), even accounting for this semi-major axis difference, the Gr{\"u}n et al. model still predicts collisional lifetimes shorter than our model. For particles in this size range, the Gr{\"u}n et al. model predicts collisional lifetimes typically less than the estimated age of the Emilkowalski cluster. This would imply that the larger particles in the partial dust band at 17$^{\circ}$ may be more collisionally evolved than our model predicts.  Nevertheless, this does not affect our argument that the partial dust band represents a much less collisionally (and dynamically) evolved distribution of particles than do the older, fully formed dust bands.

\section{RESULTS AND DISCUSSION}\label{sec:results}
\subsection{Signatures of Recent Disruptions}

\begin{figure}[htbp]
     \centering

   \begin{minipage}[b]{0.45\textwidth}
\textbf{A}\\
      \centering
      \includegraphics[bb=45 15 449 409, clip,width=\textwidth,totalheight=0.5\textheight,keepaspectratio]{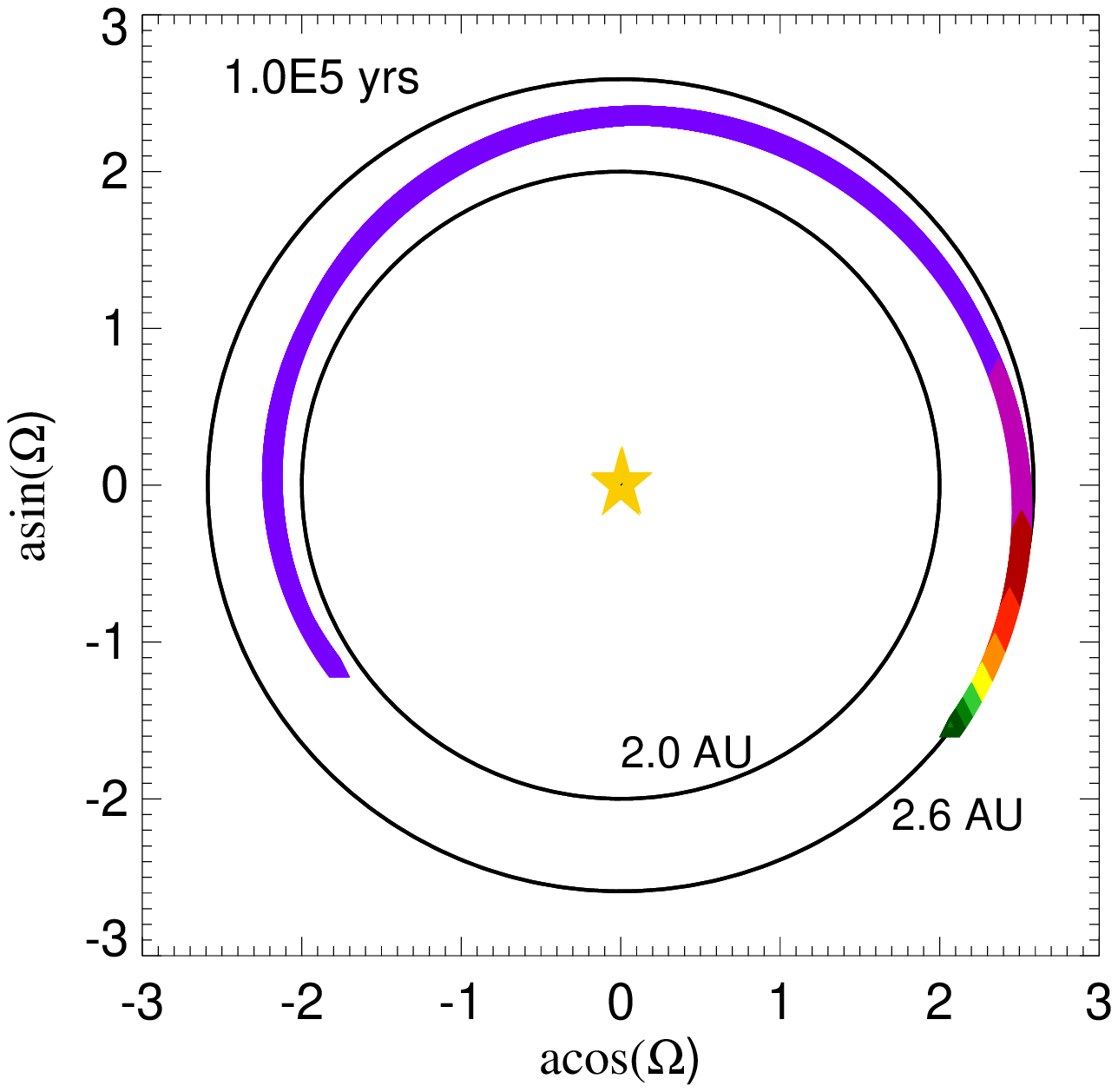}
   \end{minipage}%
   \hspace{0.01\textwidth}%
   \begin{minipage}[b]{0.45\textwidth}
\textbf{B}\\
      \centering
      \includegraphics[bb=45 15 449 409, clip,width=\textwidth,totalheight=0.5\textheight,keepaspectratio]{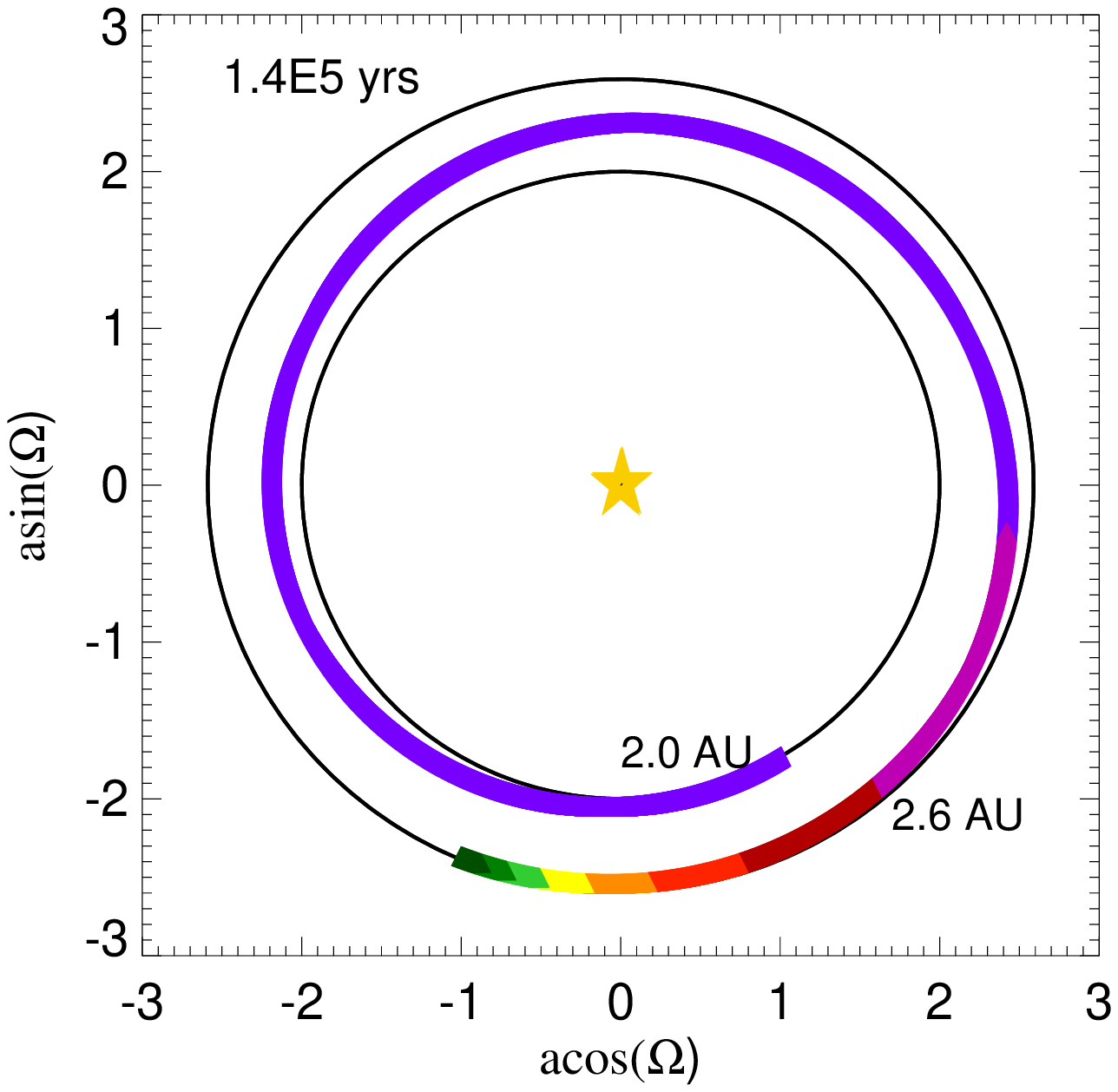}
   \end{minipage}\\[12pt]
   \begin{minipage}[b]{0.45\textwidth}
\textbf{C}\\
      \centering
      \includegraphics[bb=45 15 449 409, clip,width=\textwidth,totalheight=0.5\textheight,keepaspectratio]{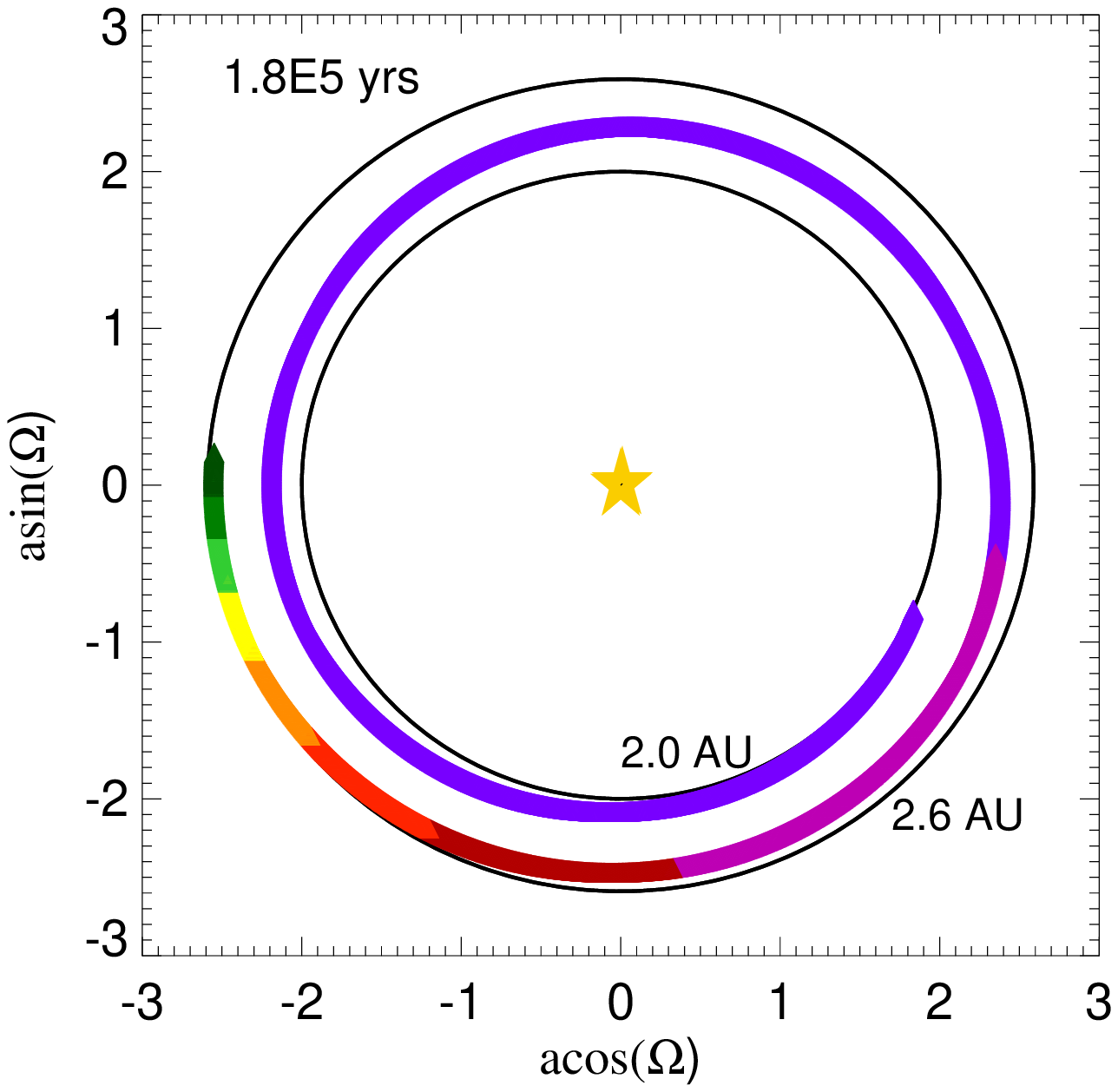}
   \end{minipage}%
   \hspace{0.01\textwidth}%
   \begin{minipage}[b]{0.45\textwidth}
\textbf{D}\\
       \centering
      \includegraphics[bb=45 15 449 409, clip,width=\textwidth,totalheight=0.5\textheight,keepaspectratio]{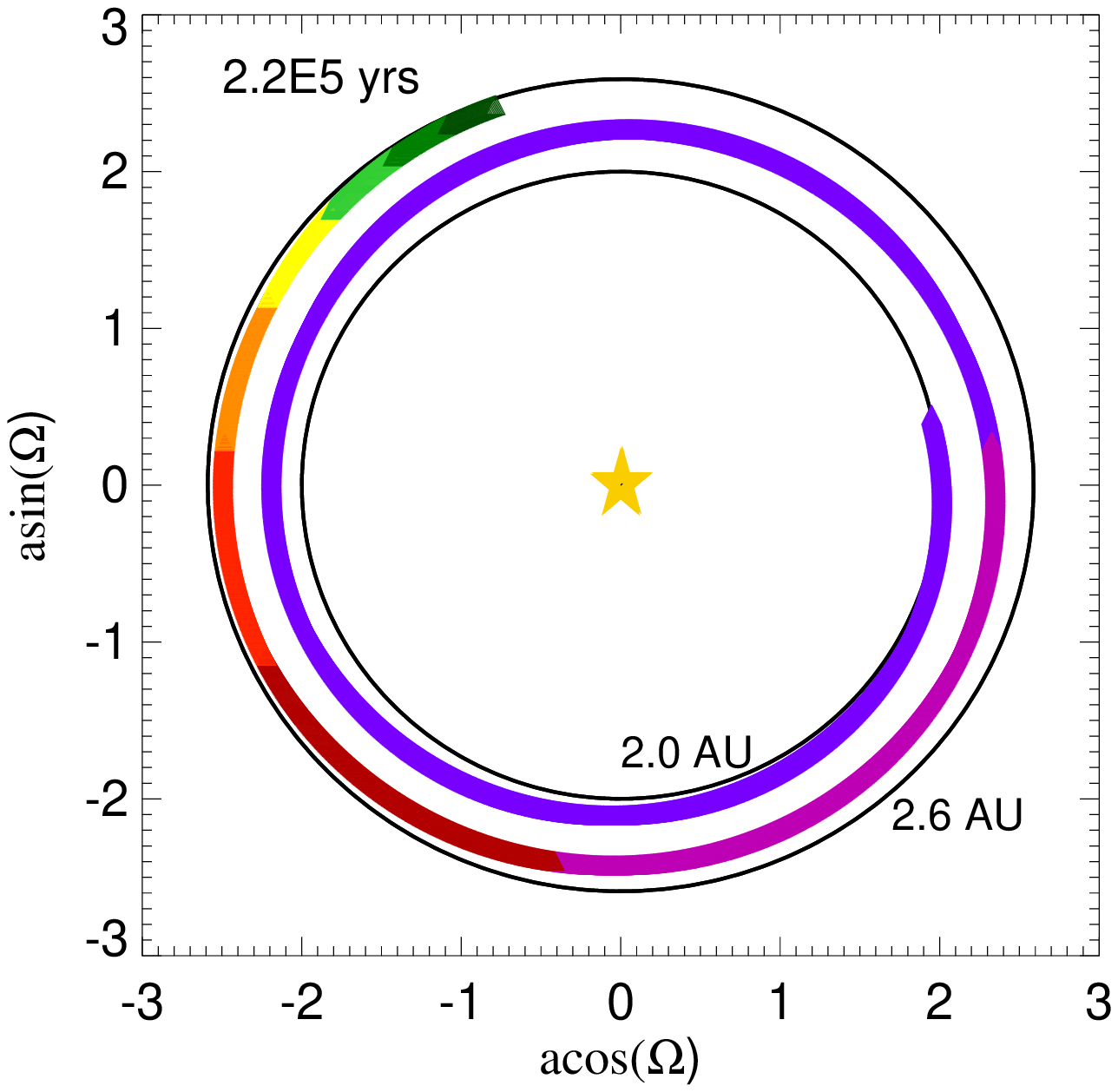}
   \end{minipage}
\caption[Nodal dispersion of particle sizes]{Time steps showing the dynamical evolution of a power-law size-distribution of dust particles released during the collisional disruption of the source body of the Emilkowalski cluster and illustrating stages in the formation of the 17$^\circ$ partial dust band. The plotted points represent the semi-major axis (a) and longitude of ascending node ($\Omega$) of a particle's orbit in a polar coordinate system and are color coded as a function of particle size (in 100 $\mu$m diameter increment bins), with the colors blue, purple, dark red, red, orange, yellow, light green, green, and dark green representing dust particles with diameters in the range 100--200 $\mu$m, 200--300 $\mu$m, 300--400 $\mu$m, 400--500 $\mu$m, 500--600 $\mu$m, 600--700 $\mu$m, 700--800 $\mu$m, 800--900 $\mu$m, and 900 $\mu$m--1 mm, respectively. The orbit of the source body of the Emilkowalski cluster is shown at its semi-major axis of 2.6 AU and the approximate inner edge of the dust bands is shown at 2 AU. At the 2.2 ky current age of the 17$^\circ$ band (bottom right panel), the nodes of the smaller particles are dispersed azimuthally around the sky, suggesting some faint dust band structure should be present at all longitudes. However, the nodes of the larger (mm to cm sized) particles remain clumped near the source and the thermal flux from these particles dominates the structure observed in the coadded IRAS observations.}
\label{fig:spirals}
\end{figure}

The semi-major axis location of particles (due to P-R drag) in this early stage of dust band formation is a function of only particle size, assuming collisions aren't yet dominating the evolution (Section \ref{sec:collisions}).  Since the precession rate of the node is a function of semi-major axis, the node location of the particle's orbit is also a function of particle size.  This results in an interesting aspect of the early stages of dust band formation, whereas the particles are distributed around the sky (in both node and semi-major axis) continuously with particle size.  The resulting spiral structure persists only for a short time in the early stages of dust band formation (between 10$^5$ and 10$^6$ years) until collisions begin to dominate.  
This is shown schematically in Figure \ref{fig:spirals} in which, as function of particle size, the location of the node and semi-major axis of the particle are shown for a series of time steps.  Once the particles are released from the parent body, the smallest particles ($\sim$100 $\mu$m shown in purple) begin to evolve away from the largest particles ($\sim$5 cm shown in green), which remain near the source.  As the semi-major axes of the smaller particles decay, the nodal precession rates slow due to their larger distance from Jupiter and its perturbing effects.  Since these smaller particles are precessing at a slower rate, they lag behind the larger particles (and the source) which results in a differential dispersion of the nodes around the sky, sorted by semi-major axis and particle size. Older, fully formed dust bands, in contrast, represent a situation where particles of all sizes exist at all semi-major axes and with randomized nodes since they have had enough time to reach this state through inter-particle collisions and orbital decay.\\

The structure of these spiral plots also serves to explain why partial dust bands contain additional information that allows us to constrain the size distribution of the material ejected from the asteroid disruption that produced them. Because the particles are effectively sorted azimuthally around the sky as a function of size, the size distribution of the dust will be revealed in the longitudinal distribution of material. Thus, for different values of the size-distribution parameter, $q$, the thermal emission intensity of the dust bands with longitude will vary. Comparison of the modeled longitudinal variation together with the observed variation of the partial band intensity in the different coadded lunes (e.g. Figure \ref{fig:variation}) allows this parameter to be constrained. Smaller values of $q$, which result in a greater proportion of large particles will yield a band with most of its intensity in the large particles nearer the source, whereas larger $q$ values which result from a population dominated by smaller particles will result in a more even distribution of band intensity around the sky.  While the pattern of intensity variation allows us to constrain the size distribution, the magnitude of the band intensity allows us to constrain the cross-sectional area of material in the band.

\subsection{Comparison of Models with Observations}\label{sec:comparetoIRAS}

By modeling different parameters describing the material producing the partial dust band at 17$^\circ$ and comparing the resulting thermal emission with the coadded IRAS data, we
have been able to constrain the cross-sectional area, the size range of particles, the size distribution, and confirm the source of the band.\\

\subsubsection{Source of the Dust Band}
We have shown through detailed modeling that the Emilkowalski cluster is the likely source of the 17$^{\circ}$ band based on its inclination, age, semi-major axis, and node.
However, we note that such a faint dust band could be produced by some yet-undiscovered asteroidal source.\\

\subsubsection{Particle Sizes}

\begin{figure}[ht]
\centerline{\includegraphics[width=\columnwidth]{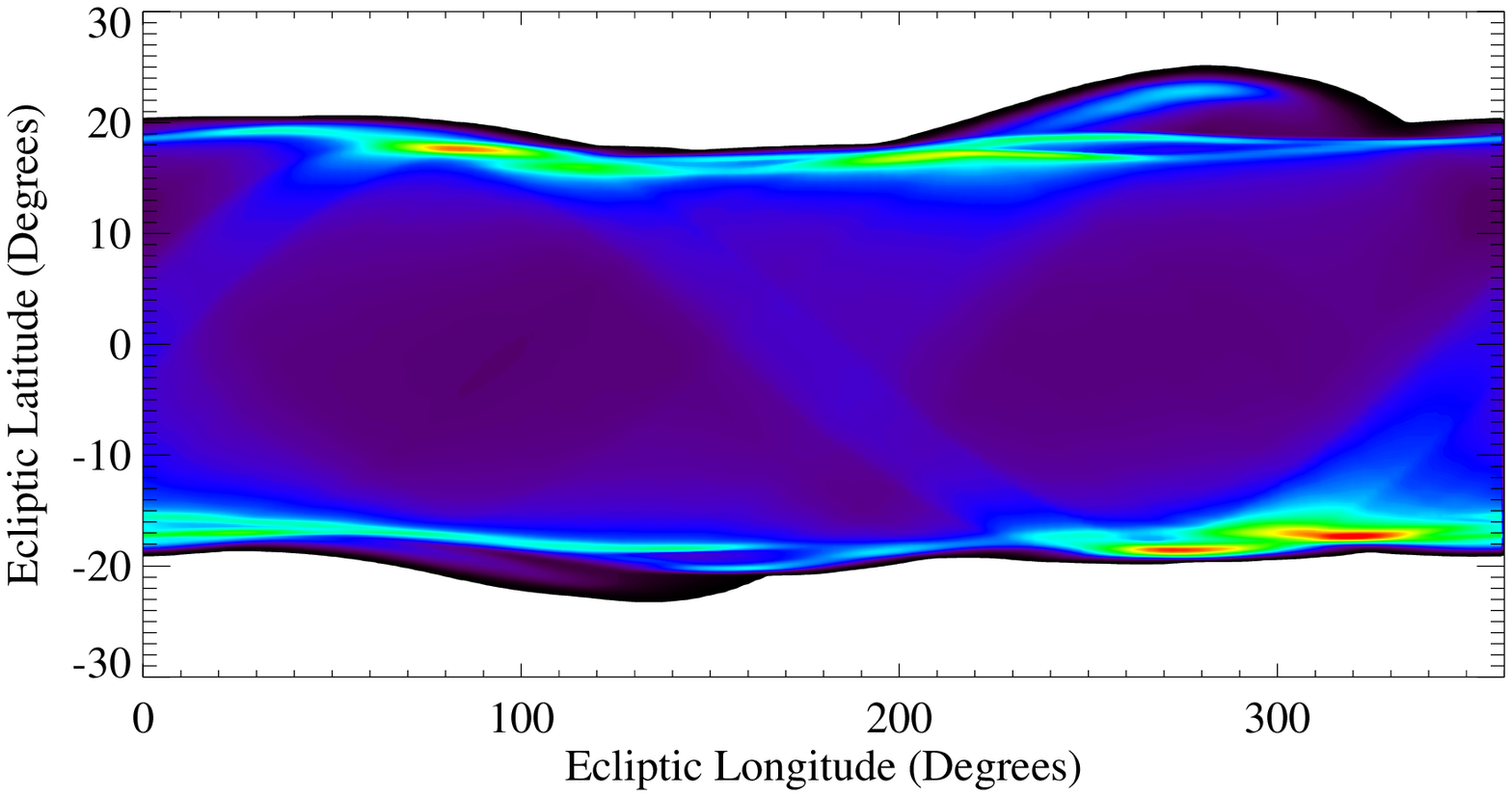}}
\caption[Example skymap]{Skymap showing the thermal emission in the 25 $\mu$m waveband of a model of the 17$^{\circ}$ partial dust band assumed to be produced by the collisional disruption of the source body of the Emilkowalski asteroid cluster. The skymap represents the intensity variation with black being the least intense and red being the most intense. The dust band can be seen as the over-density (shown in red and yellow) at the edges of the torus.}
\label{fig:skymap}
\end{figure}

Using the method described in Section \ref{sec:models}, we created
models of the dust band structure resulting from the Emilkowalski
disruption.  We used particles of diameter $\sim70~  \mu \textrm{m}-5~  \textrm{cm}$
described by a power-law size distribution ranging from $q = 1.6$ to
$q = 1.9$. Using the orbital distribution of material described by Figure
\ref{fig:spirals}, we produce skymaps of the thermal emission from the
resulting dust torus and one example of this is shown in Figure
\ref{fig:skymap}. Interestingly, the most intense thermal emission structure, described by a size-distribution parameter of
$q = 1.8$ (which is steep enough that the cross-sectional area is
dominated by small particles)  is coming from the largest particles ($>$1~mm)  in
the distribution.  This is surprising because  the
percentage of cross-sectional area coming from the larger ($>$1~mm)
diameter particles, described by the inverse power-law size
distributions used here, is small.  Additionally, these
larger particles are at a greater semi-major axis, and thus the flux
contributed from these larger particles is reduced even further. Due to the
combination of these two effects, we would not expect a significant contribution to the thermal emission structure from the largest particles.  However, for dust bands at
the age of the Emilkowalski cluster, most of the larger particles are still located near the source since they
have not yet had much time to decay in semi-major axis (under P-R and solar
wind drag) or disperse under differential nodal precession. And, even
though the smaller particles dominate both the number density and the
cross-sectional area of the material, their nodal distributions are much more dispersed.  For the current
age of the Emilkowalski disruption (Figure \ref{fig:spirals}) the smaller particles, 100--200~$\mu$m in size (shown
in purple) are distributed in node fully around the sky, while the
larger particles are clustered near the source.  So, although the
total cross-sectional area from particles $\geq$1~mm and the total
flux that they produce are both reduced compared to their smaller
counterparts, these larger particles are bunched closely together in
the longitudes of their nodes, resulting in an effectively increased
``flux-density'' compared to the smaller particles, resulting in the mm
to cm sized particles dominating the observed thermal emission structure.\\

The best fit of the models to the observations is described by a size distribution with $q>1.7$ which corresponds to a cumulative inverse power-law index larger than 2.1.  This is much steeper (e.g. more small particles) than that found for the older, central and 10$^{\circ}$ dust bands, whose particles have undergone further processing from inter-particle collisions and dynamical and radiative removal of the smaller particles. These bands are described best by a $q = 1.4$ (Grogan et al. 2001; Espy 2010), which corresponds to a cumulative inverse power-law index of 1.2.  It is also interesting to compare the size distribution found for the young, partial band to that found for the surface of asteroid Itokawa. Tsuchiyama et al. (2011) find a cumulative inverse power-law index that they describe as shallower than 2.8 and ``probably around 2.0'' for particles down to a few 10's of microns. This value for the regolith on the surface of the 350 m diameter asteroid Itokawa compares well to our findings for the dust generated in the catastrophic disruption of the 8 km diameter asteroid parent to the Emilkowalski cluster. This may imply that the dust released in the catastrophic disruption of an asteroid in this size range is dominated by the release of its surface regolith particles. For the extremely recent disruption of P/2010 A2, the tail of newly ejected dust is in the size range of mm--cm and is described by a  cumulative inverse power-law index of 2.3--2.5 (Jewitt et al. 2010; Snodgrass et al. 2010; Hainaut et al. 2011).  This is slightly steeper slope which indicates more contribution from particles in the smaller end of the size distribution, but particles below 1 mm are notably absent.  For further comparison, the size distribution found for the lunar regolith ranges from a cumulative inverse power-law of 3.1--3.3 for particles 20--500 $\mu$m (Heiken et al., 1991).\\

\subsubsection{Cross-Sectional Area of Dust Band Material}

Because the partial band is so young, modeling of this structure
allows us to constrain the cross-sectional area of dust closer to that
originally produced in the catastrophic disruption of an asteroid.
The modeling to determine the cross-sectional area in this new band is
achieved by adding an observational background, Fourier-filtering the
results, and then comparing the magnitude of the modeled and observed
bands in line-of-sight profiles.  In this way, we estimate the
cross-sectional area of material in the 17$^\circ$ dust band to be
between  10$^6$--10$^7$~km$^2$. For comparison, the cross-sectional area of dust associated with Veritas (the 10$^\circ$ band) is $O(10^9)$~km$^2$ (Espy 2010).  However, for the age of the Emilkowalski disruption, we know that semi-major axis decay due to P-R drag has resulted in the particles $\sim$70 $\mu$m and smaller in diameter being interior to the region where the dust band structure exists (exterior to approximately 2 AU).  Therefore, the constraint on the cross-sectional area of our model of the partial band, which considers only the particles contributing to the band, can be expanded to account for the $\leq$70~$\mu$m diameter particles (for a given size distribution). For a size distribution of $q = 1.8$, we find that the area of dust generated in the disruption was a factor of approximately four higher than we see today when the $\leq$70 $\mu$m diameter particles are included.  Thus, the cross-sectional area of the material associated with the partial dust band, and thus with the $\sim$10 km diameter parent body of the Emilkowalski family is on the order of 10$^7$ km$^2$.  If the volume of the parent asteroid can be used as a measure of the amount of dust released then, since the parent to the Veritas family was of order 100 km in diameter and would represent three orders of magnitude more volume than the 10 km Emilkowalski parent, we might expect that the disruption of the Veritas family initially produced on the order of 10$^{10}$ km$^2$ cross-sectional area of dust.  This implies that the cross-sectional area of material of the asteroidal component of the zodiacal cloud may have been an order of magnitude higher immediately following the disruption of the Veritas family. If the injection of dust from the Veritas disruption was originally an order of magnitude higher, then this would suggest that the asteroidal component dominated the cloud at that time. This order of magnitude increase in the dust of the zodiacal cloud following the Veritas disruption is also consistent with the work of Farley et~al.\  (2006) who measured the increase of the asteroidal dust signature found in deep sea sediments (using $^{3}$He as a proxy) and found a significant peak corresponding to the age of the Veritas family formation.  The amount of dust released in the Emilkowalski disruption would correspond to a regolith layer of $\sim$3--4 m deep on the 8 km diameter parent body asteroid's surface. 
For comparison, the $\sim$17~km diameter asteroid Eros has global fine regolith of depth varying from 10--100 m deep over the surface (Thomas et al. 2001; 2002;  Veverka et al. 2000; 2001; Robinson et al. 2001; 2002; Dombard et al. 2010; Chapman et al. 2002).  The $\sim$0.35 km diameter asteroid Itokawa has a 44 cm deep regolith composed of mm--cm size particles (Miyamoto et al. 2007).

\subsubsection{Model Discrepancies}
As we produced models of the thermal emission for comparison with observations we encountered one main discrepancy between the models and observations, and that is the amount of inclination dispersion found for the larger particles. Our models show much less dispersion on the larger particles than is seen in the coadded observations. This fact prevented us from being able to put tighter constraints on the size distribution and cross-sectional area of material present.  We investigated several different mechanisms for increasing the dispersion and found that the best explanation came from the inclusion of a greater-than-expected ejection velocity at the time of disruption. \\

\begin{figure}[ht]
\centerline{\includegraphics[width=4in]{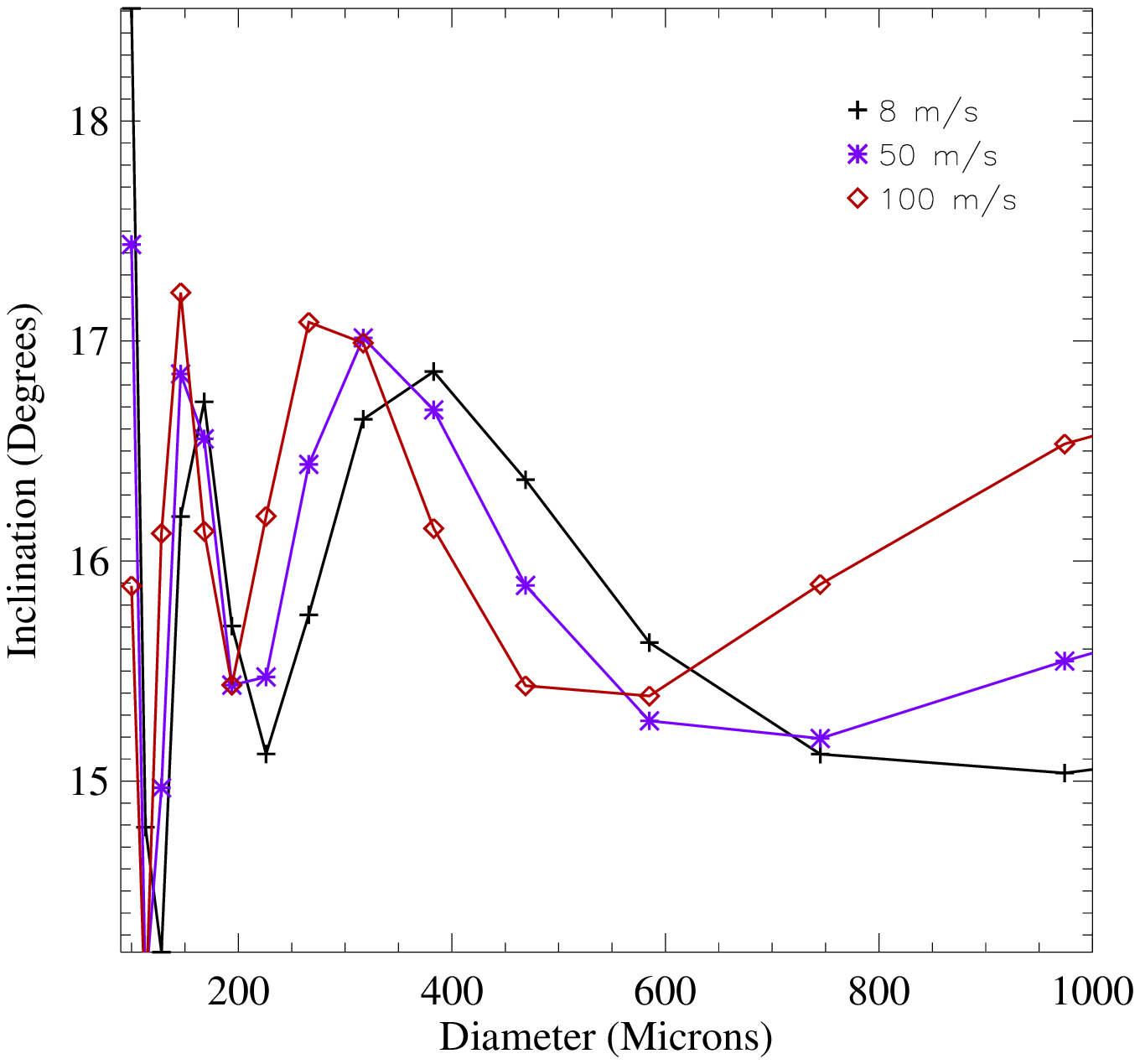}}
\caption[Effects of varying ejection velocities.]{Final orbital inclination to the ecliptic plane for Emilkowalski cluster (16.9$^{\circ}$) dust band particles (accounting for the additional dispersion accumulated during the particles' dynamical evolution) as a function of particle size is shown for three different values of the initial ejection velocity of the particles. For the larger particles, ejections at a range of velocities greater than the escape velocity ($\sim$8 m s$^{-1}$) are in agreement with the dispersion required to model the observations.}
\label{fig:veldisp}
\end{figure}

Our original model assumed a small disruption ejection velocity, one just greater than the escape velocity of the parent body, as would be expected statistically given that an asteroid is most likely to be broken up by an impactor just big enough to do so. However, even in this scenario the particles would experience a range of ejection velocities that would impart more initial dispersion (following Section \ref{section:Vel}) than we had incorporated into our original models.  Laboratory experiments and impact models show that the ejection velocity of particles is likely well above the escape speed (Housen and Holsapple 2011; Housen et al. 1979) in studies that focus mainly on crater formation, an even less energetic event than the catastrophic impacts producing asteroid families and clusters. In order to investigate how much additional orbital dispersion can be gained from higher material ejection velocities, we revisit Gauss' equations with ejection velocities of 8 m s$^{-1}$ (just over the escape velocity of the parent body), 10 m s$^{-1}$, 30 m s$^{-1}$, 50 m s$^{-1}$, 100 m s$^{-1}$, and 1 km s$^{-1}$ (which is expected only for about 1 percent of the products of the disruption of a solid parent body,e.g. Sykes and Greenberg 1986, and likely would not be expected here).  We undertook additional dynamical evolution runs to check how these initial ejection velocities would not only change the initial orbital elements, but how they would effect the final orbital elements for a particle of a given size after those particles dynamically evolve over the age of the Emilkowalski cluster. The results are given in Figure \ref{fig:veldisp}, which shows the drastic effect of even a modest (few times the escape velocity) ejection velocity.  We find that this effect provides significant inclination dispersion (on the order of a few degrees) and allows a much better match of the model to the observations.  \\

While the effects of the initial velocity dispersion seem to be the
most likely explanation for the observed inclination dispersion, the
models still don't give exactly the same same shape of the inclination
intensity distribution as is seen in the observations.  Thus, likely
multiple effects are at work, and this problem should be revisited in the future.\\

\section{CONCLUSIONS}\label{sec:conclusions}

\begin{figure}[ht]
  \centering
    \includegraphics[width=4in, scale=0.5]{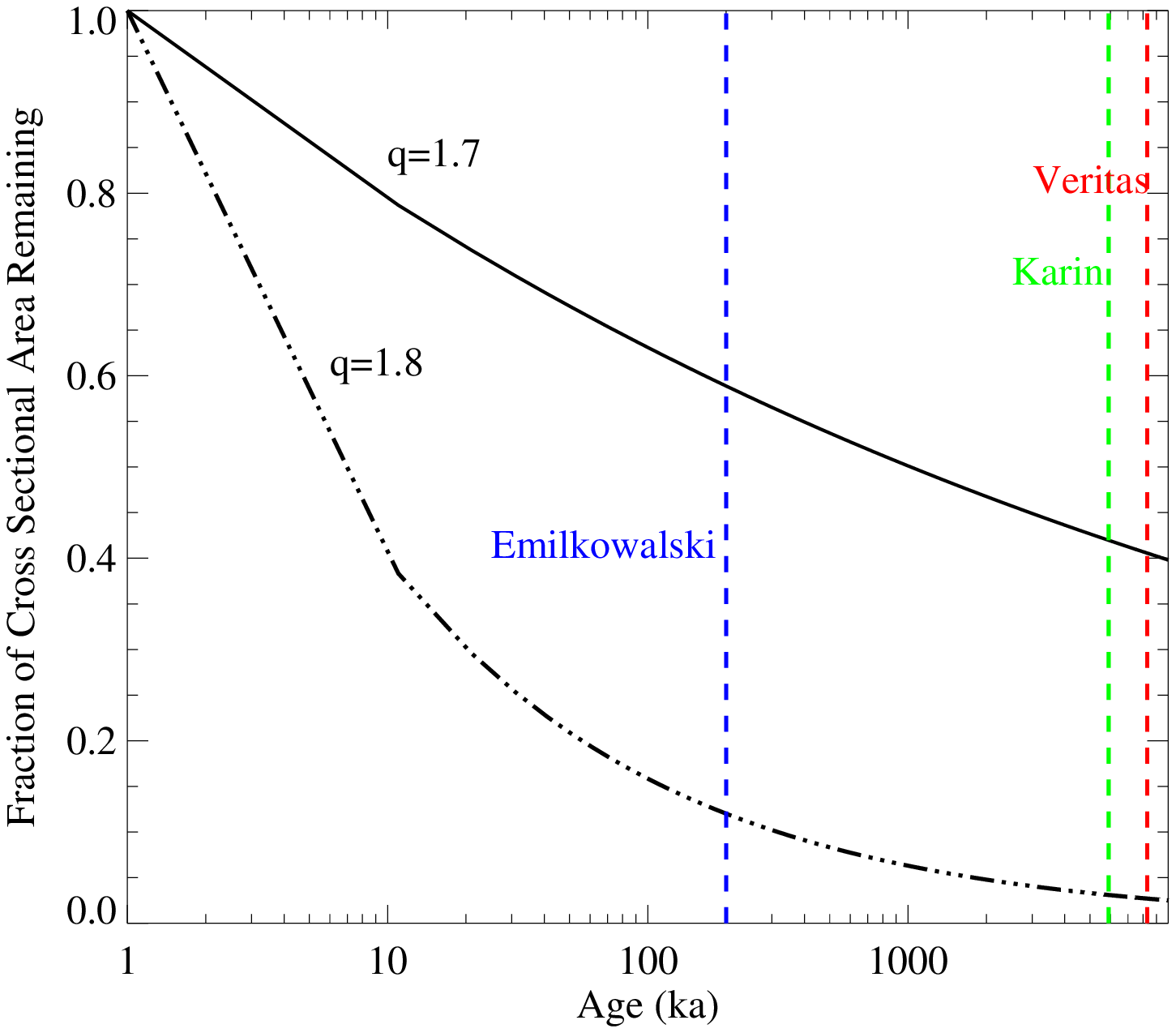}
   \caption[P-R drag decay of area with time]{ As the semi-major axis of dust particles decay under the effect of Poynting-Roberston (P-R) drag, they eventually reach inside 1 AU, where they are no longer contributing to any observations made by IRAS.  The smaller particles decay faster and are removed more quickly, thus the proportion of the area from a disruption that has been lost is a function of the size-distribution of the dust.  Higher $q$ values (which represent a steeper inverse power-law size distribution) contain more of a contribution from the smallest particles and thus decay faster than lower-$q$-value size distributions.  The curves shown here account for loss by P-R drag of the original disruption only, and not collisional loss, which would serve to further decrease the area at later ages.  The age of Emilkowalski is shown (2.2~$\pm$~0.3~ka, Nesvorn\'{y} et al. 2006b) as well as the ages of two of the older dust bands associated with Karin and Veritas (5.8~$\pm$~0.2 and 8.3~$\pm$~0.5~Ma, respectively; Nesvorn\'{y} et al. 2003).}
\label{fig:xc_v_area}
\end{figure}

	Partial dust bands, such as the one at an ecliptic latitude of $\simeq17^\circ$ studied in this paper,
        provide more constraints on their source body than do fully formed dust bands.  In a
        complete dust band, as we see for the Karin, Veritas, and
        Beagle bands, the nodes of the orbits of the dust particles
        are fully differentially precessed around the sky, erasing any
        information on the node of the parent body.  Partial dust
        bands, though, still contain enough information to constrain
        the node of the source.  Moreover, the amount of dispersion in
        the longitudes of the nodes of the dust particle orbits
        provides a constraint on the age of the disruption that
        produced them.  Older, complete dust bands are only attributed
        to a source based on their inclination, but partial bands can
        be linked to a source also constrained by the inclination,
        node, semi-major axis and age.  Additionally, in these young forming bands,
        more of the dust produced in the initial disruption is still
        present. Figure \ref{fig:xc_v_area} shows the proportion of
        the cross-sectional area from the original disruption that
        remains with age, as particles are being lost to P-R
        drag. Decay curves are shown for two different particle size
        distributions, and the ages of the young band source,
        Emilkowalski, and the older sources, Karin and Veritas, are
        marked by vertical dashed lines.  It can be seen that a greater
        proportion of the material created in the disruption of
        Emilkowalski is still present as compared to the remaining
        proportions for Karin and Veritas. Because for the older
        (Karin, Veritas, Beagle) bands, all of the original dust
        smaller than about 1 mm has already evolved inside of 1~AU
        under the effect of P-R drag, the remaining dust must have
        been created in secondary inter-particle collisions.  Thus,
        not only is much of the original dust lost, but also the
        information on the original (non-collisionally evolved)
        size distribution has also been lost.  In the early stage of
        formation represented by partial dust bands, inter-particle
        collisions have not yet begun to dominate and the size distribution of the dust is representative of that produced in the initial disruption; allowing a much tighter constraint on the size distribution and cross-sectional area of dust produced in the catastrophic disruption of the parent asteroid.\\

	Through modeling of the partial dust band, we can gain some
        information about the presence of a regolith layer on the
        parent asteroid of the Emilkowalski cluster. Analysis of the larger bodies created in the
        disruption, those that now constitute the cluster,
        allow us a reasonably good estimate of the size of the parent
        body, since most of the volume will come from the large
        members which are still present on orbits similar to that of
        the parent body.  Using these two pieces of the puzzle (the
        size of the parent body from observations of the members, and
        the cross-sectional area and size distribution of dust found in the dust band) we can try to reconstruct the structure of the parent asteroid. For example, was there a regolith layer of dust on the body and to what depth?  We have shown that the amount of dust released in the Emilkowalski disruption would correspond to a $\sim$3--4 m deep layer of regolith on the surface of the $\sim$8 km diameter parent body.  This regolith depth is on the order of 0.1\% of the diameter of the parent asteroid, which is consistent with that found for Eros (e.g. Veverka et al. 2000) and Itokawa (Miyamoto et al. 2007). \\

   	Modeling the dust in the partial bands and putting constraints
        on the regolith layer of the precursor asteroid can also help
        constrain how we expect the magnitude of the zodiacal cloud to vary
        with time (e.g. Durda \& Dermott 1997, Grogan et al. 2001, Dermott et al. 2001).  Asteroids are no
        longer thought to be just solid rocks. In a study
        of rotation rates of asteroids by Pravec \& Harris (2000),
        they find that the absence of fast rotating asteroids (less
        than 2.2 hour periods) and the tendency of the faster rotators
        to have spherical shapes is evidence that asteroids larger
        than a few hundred meters are mostly loosely bound,
        gravity-dominated, rubble pile aggregates with negligible
        tensile strength.  Britt et al. (2002) find that most asteroids have significant porosity, as indicated by bulk densities that are much less than the grain densities of their meteorite analogs. Itokawa, for example, has a bulk density of 1.9~$\pm$ 0.13~g~cm$^{-3}$ and a grain density of $\sim$3.4 g cm$^{-3}$, which correspond to a macroporosity of 39~$\pm$~6\% (Tsuchiyama et al. 2011).  If a large percentage of the asteroids are
        rubble piles, and Richardson et al. (2002) in a summary of
        evidence, also conclude that many, if not most, km-sized
        asteroids may be, then this could greatly affect how we expect
        the magnitude of the zodiacal cloud to vary with time.    A
        porous rubble pile asteroid will have absorbed numerous previous impacts, each
        contributing to the depth of the regolith layer and, thus,
        increasing the amount of dust that will be released upon the
        catastrophic disruption of the asteroid. Recent images of
        Itokawa (Fujiwara et al. 2006) and Eros (Veverka et al.
        2001) reveal how significant a regolith layer can be, from 10's of cm to 10's of m deep, respectively.\\
        
Constraining the parameters of the dust particles released in the disruption
        of a rubble pile gives insight into how the asteroidal
        fraction of the zodiacal cloud would vary with time, since
        likely a much larger amount of dust would be liberated
        from the disruption of a rubble-pile asteroid of a given size
        than from a solid body asteroid of the same size. For example,
        Dermott~et~al.~(2002) find that if a 200~km diameter asteroid
        (approximately the size of the precursor to the Eos family),
        had a regolith depth of $\sim$70 m, then the whole zodiacal
        cloud (a cross-sectional area of about $2.5~\times~10^{10}$~km$^2$)
        could have been released from its disruption.  Assuming that the depth of regolith on a body does scale roughly with the diameter of that body (as seems to be the case for Eros and Itokawa, e.g. Veverka et al. 2000; Miyamoto et al. 2007), then we might expect that Eos had a regolith depth roughly an order of magnitude higher than 70 m, suggesting that the disruption of this body could have resulted in an increase of the mass of dust in the cloud by a factor of 10, and an increase in the cross-sectional
        area of dust by a factor of $10^2$--$10^3$.  This is a more conservative estimate than that suggested by Dermott et al. (2002), who argued that a large, ancient rubble pile such as Eos could have significantly more regolith and thus the recent disruption of such a body could
        have resulted in an order of magnitude increase in dust in the cloud.  Along with the cross-sectional area of dust produced, the presence of a regolith layer could also affect how long the dust released persists since it would skew the size distribution of the dust produced in the disruption toward the smaller particles (and the P-R drag decay of the orbits is a function of particle size). \\

\noindent

	Although we only see this one partial dust band in the coadded
        IRAS data, the recent discovery of several new, young (less than 1 My old) asteroid clusters (Nesvorn\'{y} \& Vokrouhlick\'{y} 2006b) opens up
        new possible sources of faint dust bands and several more dust
        band pairs and partial bands have been postulated (Sykes
        1988; Reach et al. 1997).  How many more partial (or even
        full) dust bands are yet to be discovered?  The parent body of
        the Emilkowalski cluster, the precursor to the partial band,
        is estimated to be $\sim$10~km in diameter and based on collisional lifetimes, we expect the breakup of an asteroid this size to occur about every $10^5$ years (Bottke et al. 2005).  Since the timescale for dust band formation is $10^5$--$10^6$ years depending on location in the main belt (Sykes \& Greenberg 1986), we should therefore expect to see on the order of a few forming dust bands, due to the disruption of asteroids of this size, at any given time.  Additionally, since the lifetime of an asteroid is proportional to $\sqrt {D_{\mathrm{asteroid}}}$ (e.g. Wyatt et al. 1999), and since there are more smaller asteroids (due to the inverse power-law size distribution), then a smaller asteroid (which would produce faint dust bands) will break up even more often. \\

With its greatly increased sensitivity over previous detectors, the WISE (Wide-field Infrared Survey Explorer) dataset should hold evidence of more of these partial band pairs. WISE is an infrared satellite telescope that performed an all-sky survey, a follow-up to the 1983 IRAS survey. Since WISE should be sensitive not only to the 10 km diameter disruptions (the likely limit using the IRAS coadding method) but also even fainter and thus smaller disruptions, we should expect it to detect a minimum of a few forming dust bands, each of which can modeled using the method developed here.  Modeling the amount of cross-sectional area of material in any new bands will
allow us to refine our estimate of the total asteroidal component of
the zodiacal cloud.  Additionally, such modeling will help put constraints on the detailed temporal evolution of the
zodiacal cloud, since these faint bands come from smaller asteroids
which break up more frequently and their dust likely persists for
shorter amounts of time. Future work on this topic should investigate
how the temporal variation of the zodiacal cloud would compare to
extrasolar debris disk brightness variations, e.g. are these disks in
a steady state, or do they flare up in brightness with disruptions
within the systems (e.g. Telesco et al. 2005; Rieke et al. 2005)?
Understanding the temporal variation and the causes of such variation
of the zodiacal cloud is the first step to better understanding these
extrasolar debris disk systems.\\

We have already discussed (in Section 6.2.3) how evidence of dust band formation events can be seen in the geologic record, as Farley et al. (2006) found a signature of the the 8 Ma Veritas family formation event in deep sea sedimentary rock. For the young partial dust bands discussed in this paper, it is more appropriate to search for their geologic signature in ice cores.  While the Antarctic ice cores don't date back far enough to search for traces of the older, larger dust band creation events, they do span the relevant time frame of the disruption of the parent body of the 17$^\circ$ partial band, 2.2 ka. Wolff et al. 2006 have the longest ice core on record dating back 800 ky (EPICA/Dome C) and Fisher et al. 2013 have shown that it should be possible, in some locations, for a core to date back as far as 1.5 Ma, meaning any additional young dust bands discovered in the WISE data may also be found in these cores.\\

\subsection{Summary of Results}\label{sec:summaryl}
We find that Emilkowalski is the source of the partial dust
band, at an ecliptic latitude of 17$^\circ$, discovered in coadded IRAS data (Espy et al. 2009). The
cross-sectional area of dust released in the disruption that produced the band is on the order of 10$^7$ km$^2$; this is two orders of
magnitude less than that currently observed in the near-ecliptic and
10$^\circ$  bands (Espy 2010), albeit from a much smaller parent 
body. Scaling arguments imply that the now cometary-dust-dominated zodiacal cloud (Nesvorn\'{y} et al. 2008) has been dominated by asteroidal dust at
the epochs in the past ($\sim$8 and 6 Ma) when the larger parent bodies of the near-ecliptic and 10$^\circ$ dust bands broke up.  We find that the size distribution of dust released in the catastrophic disruption of the Emilkowalski parent body is described by a $q$ value greater than 1.7 (inverse power-law index greater than 2.1), implying cross-sectional area domination by small particles, which is a much steeper size distribution than the $q = 1.4$ found for the central and 10$^\circ$ bands (Espy 2010).  This implies that the small particles are being removed, as the dust band ages, at a faster rate (due to radiation forces) than they are being replenished (due to inter-particle collisions). The coadded observations show that the structure of the dust band is dominated by large (mm to cm sized) particles due to their closely aligned nodes. This power-law size distribution is, interestingly, in good agreement with the values describing the surface regolith of asteroid Itokawa, which is described by a cumulative inverse power-law index of less than 2.8, and likely closer to 2.0, for a regolith composed mainly of mm to cm sized particles (Tsuchiyama et al. 2011), but is shallower than the distribution determined for the particles released in the extremely recent disruption of P/2010 A2, which is described by a cumulative inverse power-law index of 2.3--2.5 (Jewitt et al. 2010; Snodgrass et al. 2010; Hainaut et al. 2011).  The inclination dispersion of the particles released in the disruption that produced the Emilkowalski cluster is more than would be expected for a low ejection velocity disruption.  We find  that the ejections velocities of dust, at disruption, of a few times the escape velocity provide a better fit of the models to the observations. For the 8~km diameter Emilkowalski parent body, when accounting for all dust particle sizes greater than the radiation pressure blowout threshold, and assuming a $q = 1.8$, the observed dust would be equivalent to a 3--4 m thick layer of regolith on the surface.

\section{ACKNOWLEDGEMENTS}
This material is based upon work supported by the National Aeronautics and Space Administration under Grant Nos. NNX12AB15G and NNX13AQ91G issued through the Science Mission Directorate ROSES Planetary Mission and Data Analysis Program and by the EU/Portuguese FCT through the WELCOME II Program. We would like to thank the referee for valuable comments that improved the quality of this paper.

\end{document}